\documentclass{article}

\PassOptionsToPackage{numbers, compress}{natbib}

\usepackage[preprint]{neurips_2023}

\usepackage[utf8]{inputenc} 
\usepackage{subcaption}
\usepackage[T1]{fontenc}    
\usepackage{amssymb}
\usepackage{url}            
\usepackage{booktabs}       
\usepackage{amsfonts}       
\usepackage{nicefrac}       
\usepackage{xcolor}         
\usepackage{graphicx}
\usepackage{multirow}
\usepackage{tikz}
\usepackage[edges]{forest}
\usepackage{natbib}
\definecolor{hidden-draw}{RGB}{20,68,106}
\definecolor{hidden-pink}{RGB}{255,245,247}
\usepackage{amsmath}
\usepackage{setspace}
\usepackage{makecell}
\usepackage{colortbl}
\usepackage{xcolor}
\usepackage{footnote}
\usepackage{enumitem}
\usepackage{pifont}
\usepackage{wasysym}
\usepackage{marvosym}

\usepackage[colorlinks=true, linkcolor=blue, urlcolor=magenta, citecolor=blue]{hyperref}
\usepackage{fancyhdr}

\usepackage{cleveref}

\definecolor{reder}{RGB}{255,0,0}

\newcommand{\good}{\textcolor{green}{\Large\textbf{\checkmark}}}
\newcommand{\bad}{\textcolor{red}{\Large\textbf{\texttimes}}} 
\newcommand{\on}{\textcolor{green}{\textbf{\checkmark}}}
\newcommand{\off}{\textcolor{red}{\textbf{\texttimes}}}

\newcommand{\qinghua}{\textsuperscript{$\ast$}}
\newcommand{\keda}{\textsuperscript{$\dagger$}}
\newcommand{\beihang}{\textsuperscript{$\ddagger$}}
\newcommand{\beida}{\textsuperscript{$\dagger\dagger$}}
\newcommand{\tianda}{\textsuperscript{$\S$}}

\title{xLLM Technical Report}

\begin{document}

\author{
Tongxuan Liu\textsuperscript{\Letter}, Tao Peng, Peijun Yang, Xiaoyang Zhao, Xiusheng Lu\qinghua, Weizhe Huang,\\
Zirui Liu\beida, Xiaoyu Chen, Zhiwei Liang, Jun Xiong, Donghe Jin, Minchao Zhang,\\
Jinrong Guo, Yingxu Deng, Xu Zhang, Xianzhe Dong\keda, Siqi Wang\beihang, Siyu Wu\beihang, Yu Wu\keda,\\
Zihan Tang\qinghua, Yuting Zeng\keda, Yanshu Wang\beida, Jinguang Liu, Meng Kang, Menxin Li,\\
Yunlong Wang, Yiming Liu\qinghua, Xiaolong Ma, Yifan Wang, Yichen Zhang\qinghua, Jinrun Yin\beida,\\
Keyang Zheng\beida, Jiawei Yin\keda, Jun Zhang\keda, Ziyue Wang\keda, Xiaobo Lin, Liangyu Liu\keda,\\
Liwei Lan\qinghua, Yang Liu\keda, Chunhua Peng, Han Liu, Songcheng Ren\beida, Xuezhu Wang\beihang,\\
Yunheng Shen\qinghua, Yi Wang, Guyue Liu\beida\textsuperscript{\Letter}, Yitao Hu\tianda\textsuperscript{\Letter}, Hui Chen\qinghua\textsuperscript{\Letter}, Tong Yang\beida\textsuperscript{\Letter},\\
Hailong Yang\beihang\textsuperscript{\Letter}, Jing Li\keda\textsuperscript{\Letter}, Guiguang Ding\qinghua\textsuperscript{\Letter}, Ke Zhang\textsuperscript{\Letter} \\
\\
\it JD.com \quad THU \qinghua \quad USTC \keda \quad BUAA \beihang \quad PKU \beida \quad TJU \tianda
}
\begingroup
\renewcommand{\thefootnote}{\Letter}
\footnotetext[0]{Corresponding authors: \{liutongxuan1, zhangke323\}@jd.com, \{guyue, yangtong\}@pku.edu.cn, hailong.yang@buaa.edu.cn, lj@ustc.edu.cn, \{dinggg, huichen\}@tsinghua.edu.cn, yitao@tju.edu.cn.}
\endgroup

\pagestyle{fancy}
\fancyhf{} 
\setlength{\headwidth}{\textwidth}
\renewcommand{\headrulewidth}{0.4pt}
\chead{xLLM Technical Report} 
\cfoot{\thepage} 

\fancypagestyle{titlepage}{
  \fancyhf{}
  \setlength{\headwidth}{\textwidth}
  \renewcommand{\headrulewidth}{0.4pt}
  \lhead{\includegraphics[height=0.9cm]{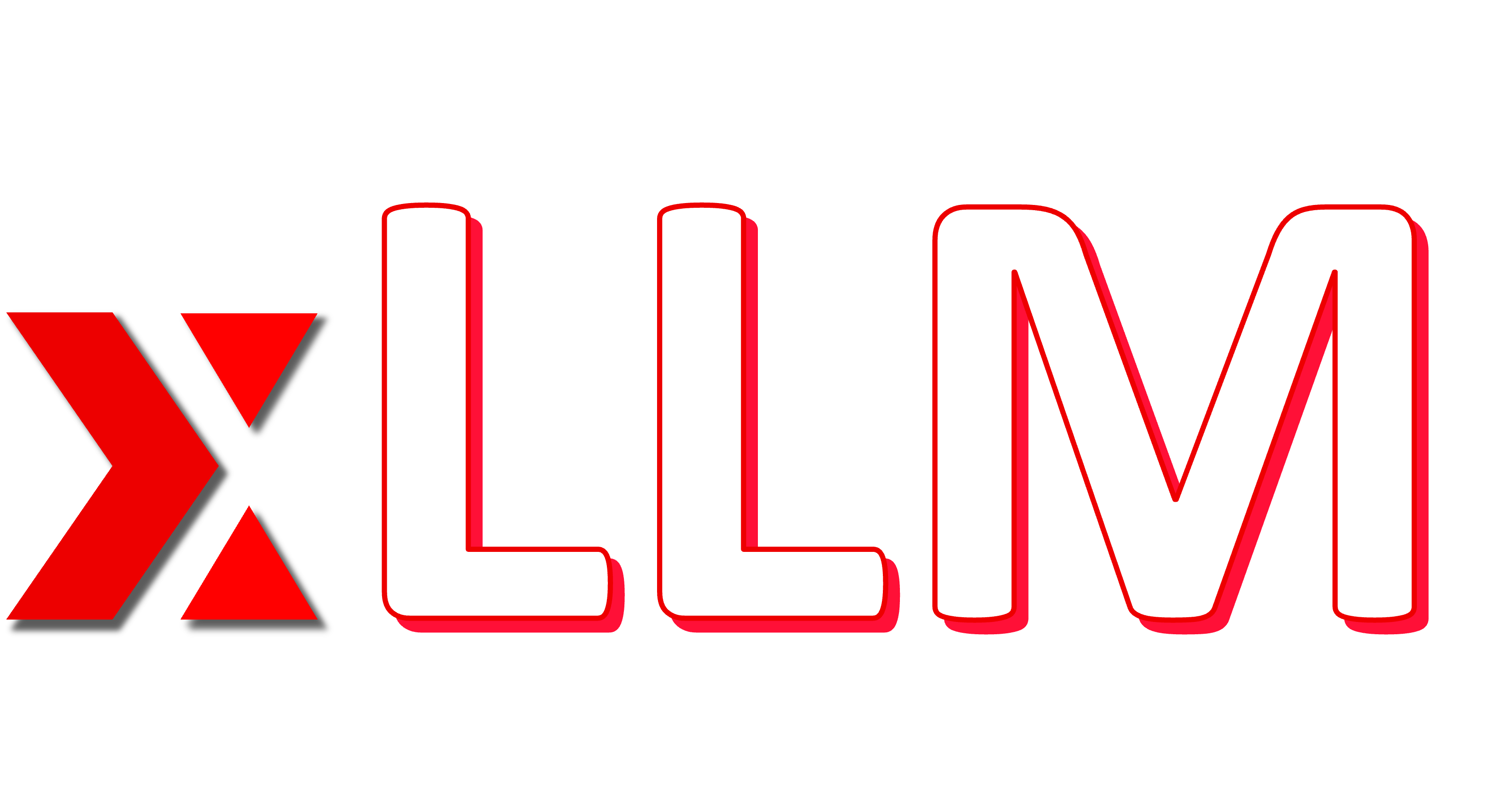}}
  
  \cfoot{\thepage}
}

\maketitle

\thispagestyle{titlepage}

\begin{abstract}
We introduce \textbf{xLLM}, an intelligent and efficient Large Language Model (LLM) inference framework designed for high-performance, large-scale enterprise-grade serving, with deep optimizations for diverse AI accelerators.
Current mainstream inference frameworks face practical challenges. On the one hand, enterprise-grade serving struggles with hybrid and dynamic workloads, strict demand for high availability of services, and distributed storage management. On the other hand, inference execution is bottlenecked by underutilized AI accelerators due to new paradigms of hardwares, model architectures and inference algorithms. 

\quad To address these challenges, xLLM builds a novel decoupled service-engine architecture. 
At the service layer, xLLM-Service features an intelligent scheduling module that efficiently processes multimodal requests and co-locates online and offline tasks through unified elastic scheduling to maximize cluster utilization. This module also relies on a workload-adaptive dynamic Prefill-Decode (PD) disaggregation policy for instance scheduling and a novel Encode-Prefill-Decode (EPD) disaggregation policy designed for multimodal inputs. 
Furthermore, it incorporates a distributed architecture to provide global KV Cache management for efficient AI accelerator memory handling and robust fault-tolerant capabilities for high availability. At the engine layer, xLLM-Engine co-optimizes system and algorithm designs to fully saturate computing resources. 
This is achieved through comprehensive multi-layer execution pipeline optimizations, including overlapping CPU scheduling with AI accelerator operations to minimize computational bubbles, employing dual-stream parallelism to overlap computation with communication, and fine-grained overlapping of various computational units to maximize hardware utilization. These are complemented by an adaptive graph mode that drastically reduces kernel launch overhead, alongside the innovative "logically contiguous, physically discrete" xTensor memory management which resolves memory allocation conflicts. xLLM-Engine also further integrates algorithmic enhancements such as optimized speculative decoding and dynamic Expert Parallel Load Balance (EPLB), collectively serving to substantially boost throughput and inference efficiency.

%
%
\quad Extensive evaluations demonstrate that xLLM delivers significantly superior performance and resource efficiency. Under identical TPOT constraints, xLLM achieves throughput up to 1.7$\times$ that of MindIE and 2.2$\times$ that of vLLM-Ascend with Qwen-series models, while maintaining an average throughput of 1.7$\times$ that of MindIE with Deepseek-series models.
%
We have deployed xLLM in production to support a range of core business scenarios at JD.com, covering areas including LLM, Multimodal Large Language Model (MLLM), and generative recommendation. These applications encompass the JingYan AI chatbot, marketing recommendations, product understanding, customer service assistants, and more.
xLLM framework is publicly available at \url{https://github.com/jd-opensource/xllm} and \url{https://github.com/jd-opensource/xllm-service}.
\end{abstract}
\clearpage

\tableofcontents
\clearpage

\section{Introduction}

In recent years, large language models (LLMs) with parameters ranging from billions to trillions (GPT~\cite{gpt}, Claude~\cite{claude}, DeepSeek~\cite{deepseek}, LLaMA~\cite{llama}, etc.) have achieved breakthrough progress in the fields of natural language processing and multimodal interaction, which drives an urgent demand in the industry for efficient inference engines and service systems. These models are being rapidly deployed in core business scenarios such as intelligent customer service~\cite{customer-service}, real-time recommendation~\cite{recommendation}, and content generation~\cite{content-generation}. However, how to reduce the cost of model inference and improve computing efficiency remains a key challenge for large-scale commercial serving. 

Current mainstream LLM inference frameworks (vLLM~\cite{vllm}, SGLang~\cite{sglang}, TensorRT-LLM~\cite{tensorrt}, etc.) face four key challenges in enterprise-level serving scenarios: \textit{First}, in an inference cluster with hybrid deployment, online inference requests exhibit significant tidal characteristics~\cite{tidal-character, fapes}. Current scheduling systems fail to meet the service level objective (SLO) for online services while fully leveraging idle periods of online services to increase the throughput of offline tasks. \textit{Second}, existing Prefill-Decode (PD) disaggregation architecture~\cite{distserve, dynamo, tetriinfer} assumes static resource allocation for the two phases, which cannot adapt to the dynamically changing request loads (i.e., input/output lengths fluctuate) in real-world applications, resulting in low AI accelerator utilization and increased SLO violation risks. \textit{Third}, there is a lack of strategy to efficiently service multimodal requests (i.e., image, voice and text input~\cite{multimodal1, multimodal2}), including parallel processing for the encode phase and fine-grained resource allocation accordingly. \textit{Fourth}, as the scale of the inference cluster increases, ensuring fast fault detection and service recovery for nodes or instances is critical to maintaining the stability of inference services.

The evolving computing paradigms also present significant performance challenges on existing LLM inference engines: \textit{First}, they struggle to fully utilize the computing units of modern AI accelerators~\cite{npu, tpu}. \textit{Second}, the All-to-All communication overhead~\cite{fastermoe} and expert parallel (EP) load imbalance in the Mixture of Experts (MoE) architecture~\cite{gshard,expertflow} restrict the scalability of the system. \textit{Third}, as the model context window continues to expand, efficient KV Cache management becomes critical to inference performance~\cite{qin2025mooncake}. \textit{Forth}, Due to the unpredictable nature of inference requests, conventional static scheduling and operator strategies face difficulties in effectively balancing workloads across computing units in Data Parallelism (DP).

To address above challenges , we propose \textbf{xLLM}, an efficient and intelligent LLM  inference framework featuring a service-engine decoupled design. xLLM achieves efficient support for enterprise-level inference through the following innovations: at the \textit{service layer}, xLLM has made groundbreaking achievements in 1) unified elastic scheduling for online/offline requests, 2) workload-adaptive dynamic PD disaggregated architecture, 3) novel Encode-Prefill-Decode (EPD) disaggregation for multimodal requests, and 4) a distributed cache management and fault-tolerance framework; at the \textit{engine layer}, xLLM enhances the resource efficiency across the full-stack of ``communication-computation-storage'', including 1) a multi-layer pipeline execution mechanism, 2) efficient computing and memory optimization, and 3) intelligent algorithm designs.

Specifically, \textbf{xLLM-Service} schedules online requests with preemptive execution and offline requests in a best effort manner, to maximize resource utilization while strictly ensuring SLOs for online services. To address the inherent limitations of the static PD configuration, xLLM incorporates an adaptive scheduler to dynamically adjust the proportion of PD instances for each request and supports fast role reversal for instances, by monitoring key metrics such as Time to First Token (TTFT) and Time Per Output Token (TPOT)~\cite{infer-metric}. For multimodal visual requests, xLLM automatically selects the optimal EPD phase-disaggregation strategy based on pre-profiling, for the best performance trade-off between throughput and latency. To handle failures such as hardware problems, network faults, or software errors~\cite{node-fault}, xLLM collects node error information, evaluates KV recomputation or migration costs of interrupted requests, and makes optimal global rescheduling decisions.
Across multiple instances, xLLM supports KV offloading and routing migration within a hybrid storage architecture, improving KV storage capacity and cache hit rates.

\textbf{xLLM-Engine} employs a multi-layer execution pipeline that incorporates hardware-specific optimizations: (i) at the framework scheduling layer, it implements asynchronous CPU-accelerator scheduling to minimize computational idle time; (ii) at the model graph layer, it utilizes dual-stream micro-batch parallelism to overlap computation with communication; (iii) at the operator level, it achieves kernel computation and memory access overlapping. For computational efficiency, xLLM-Engine automatically fuses small kernels into a unified computation graph and dispatches them to the accelerator in a single operation, significantly reducing kernel launch overhead. Regarding memory management efficiency, xLLM-Engine introduces the xTensor memory management scheme, which employs a ``logically contiguous, physically discrete"" KV cache storage structure that resolves the conflict between memory contiguity requirements and dynamic allocation needs. To further enhance performance, xLLM-Engine incorporates an adaptive speculative decoding mechanism, a redundant expert-based load balancing algorithm for EP, and a hierarchical load balancing algorithm for DP. Additionally, xLLM-Engine provides scenario-specific optimizations, such as for generative recommendation -- one of JD.com's core businesses -- where it achieves 23\% performance improvement through host-kernel operation overlapping.

\begin{figure}[t]
  \centering
  \includegraphics[width=\textwidth]{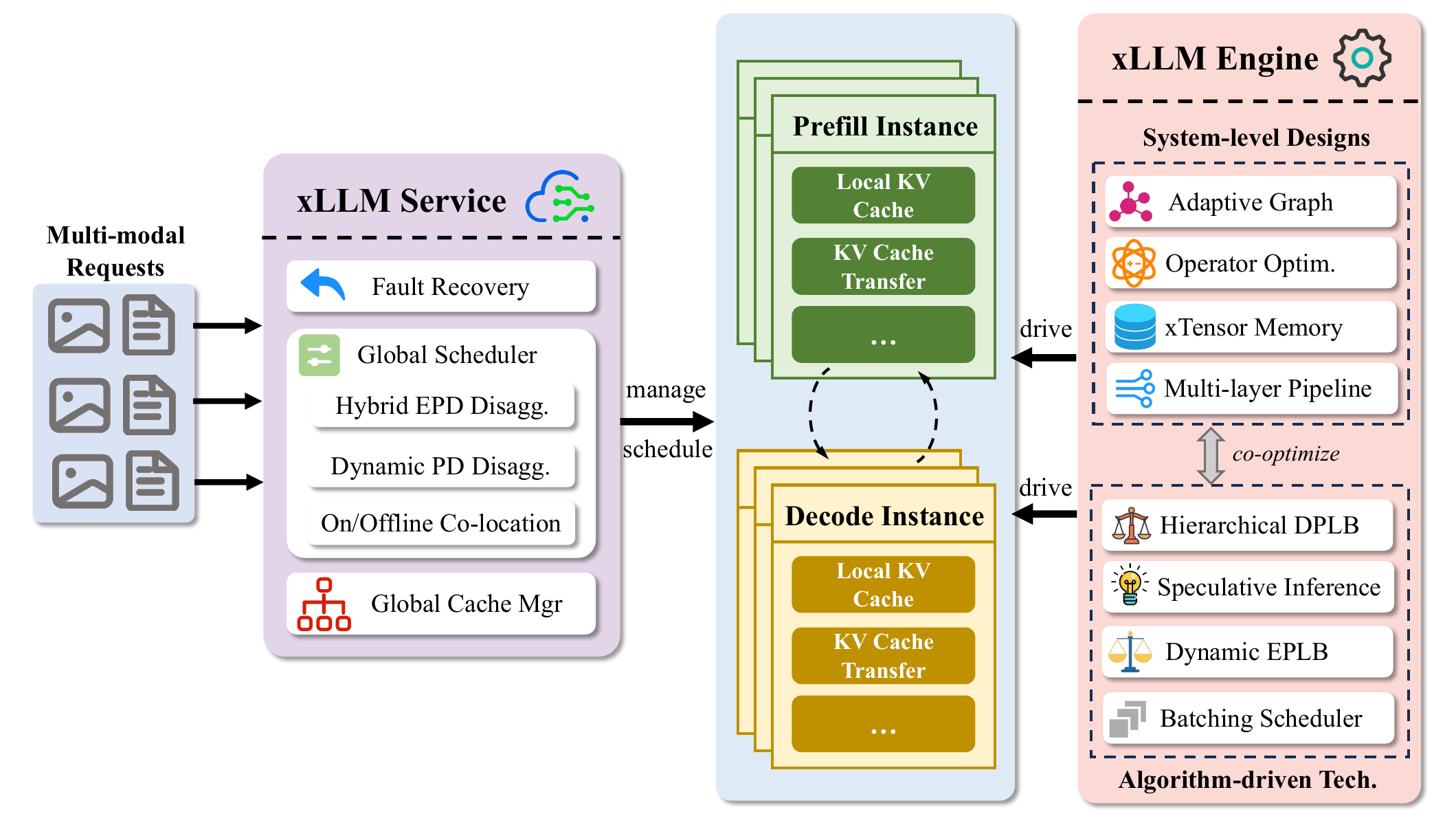} 
  \caption{Overview of xLLM's capabilities, with decoupled service-engine architecture.}
  \label{fig:capability}
\end{figure}

In summary, we make the following contributions when developing xLLM:

\textbf{xLLM Intelligent Service Capabilities}
\begin{itemize}[leftmargin=*]
\item We design a unified scheduling algorithm for online/offline workloads (\textsection\ref{sec:unified}). 
\item We implement a workload-adaptive PD disaggregation policy to address scenarios with rapidly changing traffic load and request input/output lengths (\textsection\ref{sec:pd-adapter}).
\item We propose a hybrid EPD disaggregation policy for multimodal requests, achieving intelligent resource allocation across different phases (\textsection\ref{sec:epd-planner}).
\item We leverage multi-level KV Cache management and global KV Cache routing strategy to expand KV Cache capacity and improve cache hit rates (\textsection\ref{sec:cache-mgr}).
\item We design a multi-node fault tolerance architecture to ensure high service availability (\textsection\ref{sec:fault-tolerance}).
\end{itemize}

\textbf{xLLM Intelligent Engine Capabilities}
\begin{itemize}[leftmargin=*]
    \item We achieve intelligent and efficient inference, through hardware-software co-design to improve hardware computing efficiency, including pipeline execution (\textsection\ref{sec:pipeline}), graph optimization (\textsection\ref{sec:acl-graph}), and memory optimization (\textsection\ref{sec:xtensor}).
    \item We enhance inference performance through algorithmic optimizations (\textsection\ref{sec:algo}), including optimized speculative decoding (\textsection\ref{sec:mtp}), dynamic expert parallel load balance (\textsection\ref{sec:eplb}), and hierarchical data parallel load balance (\textsection\ref{sec:dp-balance}).
    \item We optimize online inference for generative recommendation scenarios (\textsection\ref{sec:generative}).
\end{itemize}


\section{System Overview}
The overall architecture of xLLM is depicted in Figure~\ref{fig:capability}. Upon request arrival, xLLM-Service performs intelligent scheduling to distribute each request to one of three elastic instance pools and manages instance migration across these pools during runtime. xLLM-Engine then drives efficient request inference by orchestrating system- and algorithm-level optimizations.

\subsection{xLLM-Service}
\label{sec:overview-service}

\begin{figure}[t]
  \centering
  \includegraphics[width=0.75\textwidth]{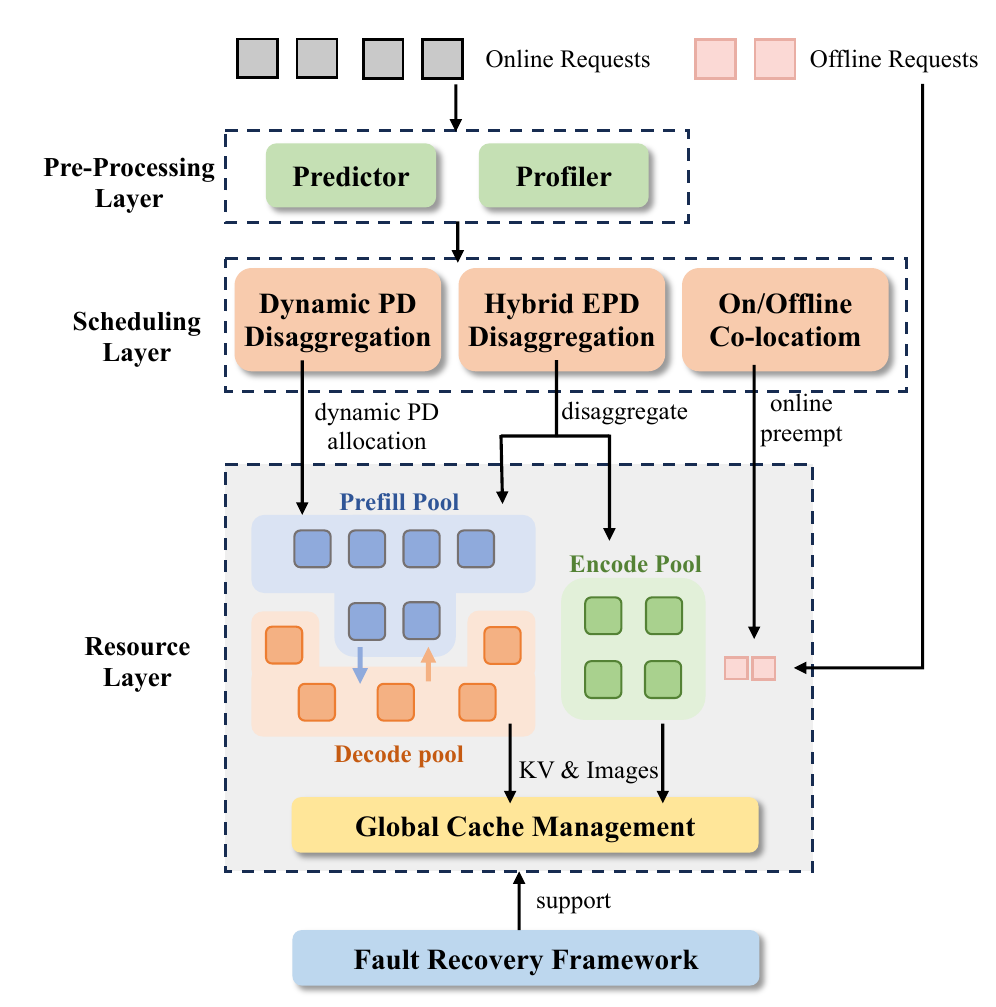} 
  \caption{System workflow of xLLM-Service.}
  \label{fig:service-workflow}
\end{figure}

We describe the workflow of xLLM-Service in Figure~\ref{fig:service-workflow}. The system comprises three primary layers: 1) a preprocessing layer consisting predictor and profiler,  2) a scheduling layer integrating three policies (\textit{Dynamic PD Disaggregation Policy}, \textit{Hybrid EPD Disaggregation Policy}, and \textit{Online-Offline Co-location Policy}), and 3) a resource layer consisting of three heterogeneous instance pools. Specifically, the \textit{Dynamic PD Disaggregation Policy} leverages information from the predictor to dynamically convert PD instances based on the workload status of the two instance types. The \textit{Hybrid EPD Disaggregation Policy} utilizes the profiler to determine the optimal EPD disaggregation strategy for multimodal requests. Concurrently, the \textit{Online-Offline Co-location Policy} dispatches requests according to their online or offline attributes. KV and image caches are offloaded and routed among distributed instances, while the fault recovery framework ensures the high availability of the entire service. 

\paragraph{Elastic Instance Pools.} Instances in a cluster are partitioned into three elastic pools: 

$\triangleright$ \, Prefill Instance Pool: This pool handles the prefill phase of text requests.

$\triangleright$ \, Decode Instance Pool: This pool handles the decode phase of text requests.

$\triangleright$ \, Encode Instance Pool: This pool handles the encode phase of multimodal requests. 

We design the instances in Prefill and Decode pools as stateless, i.e., instances do not require physical migration between instance pools. Instead, they achieve flexible switching between Prefill-Decode roles based on the type of request being processed. 

\paragraph{Request Preprocessing.} xLLM-Service implements unified resource management for both online and offline requests. Online requests are submitted in a preemptive and deadline-prioritized manner, and are co-deployed with best-effort offline tasks within the shared resource pool. During the runtime of an offline request, the \textit{Unified-Scheduler} dynamically scales it up/down based on the tidal traffic characteristics of online requests. When an online request is submitted, it first passes through the pre-processing layer, which consists of two modules: 

$\triangleright$ \, TTFT Predictor: A TTFT prediction model built for text requests. It evaluates SLO fulfillment by analyzing queueing delays from each prefill instance queue and request input lengths, thereby guiding instance allocation for the \textit{Dynamic PD Disaggregation Policy} in the scheduling layer.

$\triangleright$ \, EPD Profiler: A profiler for multimodal requests that uses binary search to identify optimal deployment configurations: (1) EPD separation strategy, choosing from three approaches: EP-D (i.e., aggregated execution of Encode and Prefill phases, with Decode phase executed separately), ED-P, or E-P-D; (2) The maximum batch size for the Encoder phase; (3) The maximum number of tokens for Prefill/Decode's inputs. The \textit{Hybrid EPD Disaggregation Policy} in the scheduling layer will use the optimal configuration determined for phase disaggregation and task dispatching.

\paragraph{Intelligent Scheduling.} The intelligent scheduling layer adjusts resource allocation for requests across their full lifecycles. It contains three major scheduling policies designed for various scenarios:

$\triangleright$ \, Online-Offline Co-location Policy: This policy implements a preemptive scheduler for managing online and offline requests. When the load of online requests reaches the peak, they preempt some offline requests on PD instances. To surrender resources to offline requests, when the load on P instances decreases (usually earlier than on D instances), P instances continue to process offline prefill requests and also migrate decode offline requests from D instances to P instances.

$\triangleright$ \, Dynamic PD Disaggregation Policy: This adaptive scheduling policy is responsible for dynamically managing P and D instance allocation. It intelligently assigns requests to suitable instances using a heuristic algorithm guided by the TTFT Predictor. Furthermore, it implements a feedback mechanism by continuously collecting performance data from computing instances, enabling runtime monitoring and adjustment of allocation decisions to maintain system efficiency.

$\triangleright$ \, Hybrid EPD Disaggregation Policy: This multimodal policy executes the three-phase disaggregation based on strategies searched by the EPD Profiler. For the EP-D disaggregation, the fused EP phase executes in the P instance pool; for the ED-P disaggregation, the fused ED phase executes in the D instance pool; for the E-P-D disaggregation, the three phases execute separately in the three instance pools. This deployment also enables multimodal requests to benefit from the adjustment of \textit{Dynamic PD Disaggregation Policy}.

\paragraph{KV-centric Storage Architecture.} The instance storage employs a hybrid architecture (HBM-DRAM-SSD) to cache KV values and image tokens. At the global level, xLLM takes the idea from Mooncake Store~\cite{qin2025mooncake} and extends it to domestic accelerators with specific optimizations. Routing and reuse of caches across instances are determined by embedded intelligent routing strategies. 

\paragraph{Efficient Fault-tolerant.} The fault recovery framework of xLLM-Service supports fault detection and fast recovery of instances from the three elastic pools (E, P, D). For requests on failed instances, the architecture manages the migration of image caches between instances, and automatically decides the optimal KV recomputation or migration strategy for handling affected KV caches.

\subsection{xLLM-Engine}
\label{sec:overview-engine}

During request execution, the xLLM-Engine layer provides intelligent computing capabilities. We achieve various joint inference accelerations between the computing system layer and the algorithm-driven layer in the following ways: 

\textbf{Computing System Layer}

$\triangleright$ \, Multi-layer Pipeline Execution Engine: In the framework layer, CPU tasks are scheduled asynchronously to create a pipeline with inference computation, reducing computational bubbles; in the model graph layer, a single batch is split to form a pipeline between two micro-batches, effectively overlapping computation and communication; in the operator kernel layer, operations are pipelined across different computational units, enabling computation and memory access to overlap.

$\triangleright$ \, Graph Optimization for Dynamic Inputs: Small kernels in the decoding stage are fused into a single computational graph. To handle variable sequence lengths and batch sizes, we parameterize the input dimensions, and employ a multi-graph caching scheme to reduce compilation overhead.

$\triangleright$ \, xTensor Memory Management: It uses a ``logically continuous, physically discrete'' KV storage structure. It allocates physical memory space on-demand during token generation for each request, while asynchronously predicting and intelligently mapping the physical pages required for the next token. When a request completes, the existing physical memory will be reused to execute the next request.

\textbf{Algorithm-Driven Layer}

$\triangleright$ \, Speculative Decoding: xLLM-Engine incorporates an optimized speculative inference algorithm that generates multiple tokens at once to boost throughput~\cite{mtp}. It further optimizes the computing architecture through ways like asynchronous CPU processing and reduced data transfer.

$\triangleright$ \, EP Load Balance: For MoE models, xLLM implements expert weight updates based on historical expert load statistics, enabling effective dynamic load balancing during inference.

$\triangleright$ \, DP Load Balance: For data parallel (DP) deployment, xLLM achieves fine-grained load balance by kvcache-aware instance allocation, inter-DP requests migration, and intra-DP computing units allocation.

We will elaborate on the detailed design of xLLM-Service in \textsection\ref{sec:service} and outline the optimizations within the xLLM-Engine in \textsection\ref{sec:engine}. A comprehensive evaluation of xLLM will be presented in \textsection\ref{sec:evaluation}.
\section{xLLM-Service Designs}
\label{sec:service}

\subsection{Online-Offline Co-location Scheduler Policy}
\label{sec:unified}

\begin{figure}[t]
  \centering
  \includegraphics[width=0.9\textwidth]{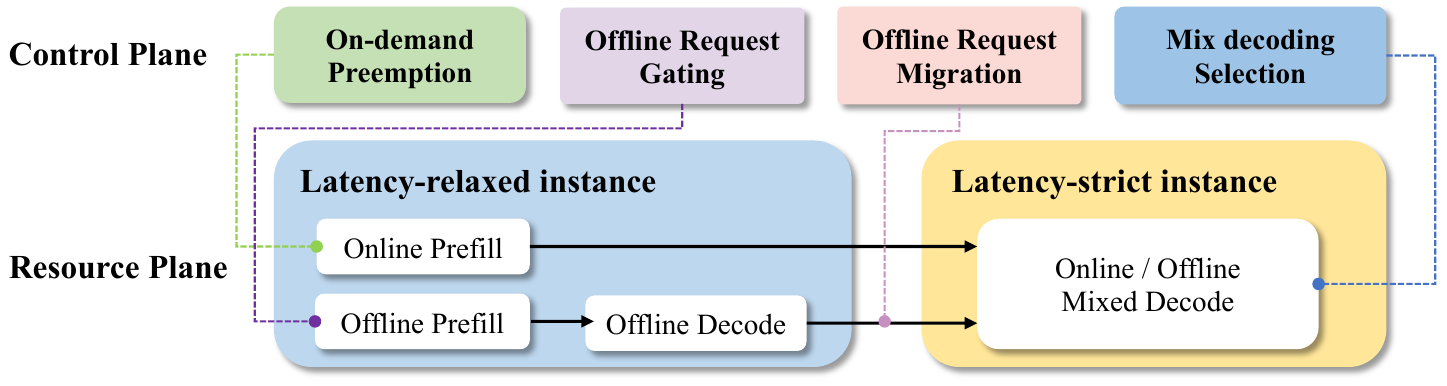} 
  \caption{Overview design of Online-Offline Co-location Scheduler Policy.}
  \label{fig:unified-scheduler}
\end{figure}

\paragraph{Online/Offline Request Characteristic.} LLM services can be categorized into two types based on their service modes: online and offline requests. Online requests, including those from chatbots~\cite{gpt, claude, deepseek-r1}, code completion~\cite{qwen2.5-coder, starcoder, copilot}, and recommendation systems~\cite{onerec, gr-llms}, constitute latency-sensitive workloads. 
These services must respond immediately upon request arrival, often returning each generated token in real time via streaming output. Consequently, they impose strict SLO requirements on TTFT or TPOT to ensure a satisfactory user experience. In contrast, offline services, such as document analysis~\cite{bowen2009document} and intelligent data annotation~\cite{karim2025transforming}, are non-real-time workloads with minimal latency constraints and thus no stringent SLO requirements.

Furthermore, we observe that the request traffic for online services typically exhibits significant volatility, including tidal variations at hourly or daily scales and sudden bursts at minute-level intervals~\cite{tidal-character,fapes}. While cluster auto-scaling~\cite{auto-scaling} could theoretically mitigate resource underutilization caused by tidal patterns, the slow cold-start latency of instances—involving model loading and complex initialization—renders it ineffective for responding to rapid traffic spikes~\cite{deepboot}, increasing the risk of service-level agreement violations.

\paragraph{Online/Offline Request Co-location Deployment.} To address these practical challenges, we adopt a hybrid deployment strategy that co-locates online and offline requests. In such a system, offline requests can utilize idle resources during off-peak periods of online traffic. Conversely, during online traffic peaks, offline tasks can be preempted, as they are not bound by strict SLOs. This approach significantly enhances aggregate resource utilization and mitigates idleness during traffic troughs. Although some recent work~\cite{sun2025hygen, wang2025echo, borui2025efficient} also attempts to co-locate offline requests, they have not explored the multi-instance scenario, especially for PD disaggregation.

Actually, the PD disaggregation architecture has demonstrated superior latency performance and is increasingly becoming a mainstream design paradigm in industry~\cite{distserve, qin2025mooncake}. However, directly applying online-offline co-location to PD-disaggregated systems introduces critical PD load imbalance issues: (1) such systems require the load ratio between prefill and decoding stages to align with their respective resource allocation ratios; otherwise, one stage may become a bottleneck, causing blocking or resource underutilization in the other. (2) Furthermore, as these PD load variations exhibit high volatility and burstiness similar to online traffic patterns, existing PD disaggregation techniques struggle to effectively address this challenge.

\paragraph{Latency-Constrained Decoupled Architecture.} We rethink the design of PD disaggregation architecture under online-offline co-location deployment. The latency advantage of PD disaggregation essentially stems from the separation of latency constraints: the decoding phase is highly sensitive to per-step latency and cannot be blocked by long-duration operations, thus necessitating decoupling from the prefill phase.
Inspired by this insight, we propose a latency-constrained decoupled architecture, as illstrated in Figure~\ref{fig:unified-scheduler}. This design regards cluster resources as two pools: a \textit{latency-relaxed} pool (corresponding to original Prefill instances) and a \textit{latency-strict} pool (corresponding to original Decode instances). All tasks are then reassigned to one of the two resource pools based on their inherent latency characteristics and requirements. Within this architecture, the decoding phase of offline requests can be executed in either resource pool. This flexibility allows us to dynamically adjust the load ratio between the two pools, thereby maximizing the overall resource utilization of the cluster. 

However, it also introduces another two challenges: (1) Complex Scheduling Space: since the decoding of offline requests can be performed on either type of instance, how to leverage this flexibility to design new intelligent scheduling remains unknown. (2) Strict SLO Guarantee: the execution of offline requests consumes resources, their prefill phase may block newly arrived online requests, and their decode phase on \textit{latency-strict} nodes may slow down the overall response speed. Both scenarios can compromise the SLO satisfaction of online requests. 

\paragraph{Solution 1 - Performance Bottleneck Analysis.} We construct an LLM inference performance model based on the Roofline Model~\cite{roofline} and online factor learning. This model is designed to predict the latency, computation utilization, and memory utilization of both the prefill and decode phases. Since decoding operations executed on \textit{latency-strict} instances typically account for a large proportion of workloads and are highly performance-sensitive, we set balancing computational and memory resources as the optimization objective. By analyzing performance bottlenecks through this model, we can select more appropriate offline requests to merge into decoding batches, thereby improving resource utilization efficiency.

\paragraph{Solution 2 - Efficient Preemption Mechanism.} To strictly ensure that the SLO of online requests remains unaffected, we introduce a preemption mechanism that allows online requests to preempt offline requests. For offline prefill tasks running on \textit{latency-relax} nodes, we propose a model execution interruption technique, which enables preemption within an acceptable latency range without incurring additional model maintenance overhead. For decoding tasks running on \textit{latency-strict} nodes, we leverage the performance model to dynamically select requests for decoding batching, ensuring that decoding latency always meets SLO constraints.

\subsection{Dynamic PD Disaggregation Scheduler Policy}
\label{sec:pd-adapter}

\begin{figure}[t]
  \centering
  \includegraphics[width=0.7\textwidth]{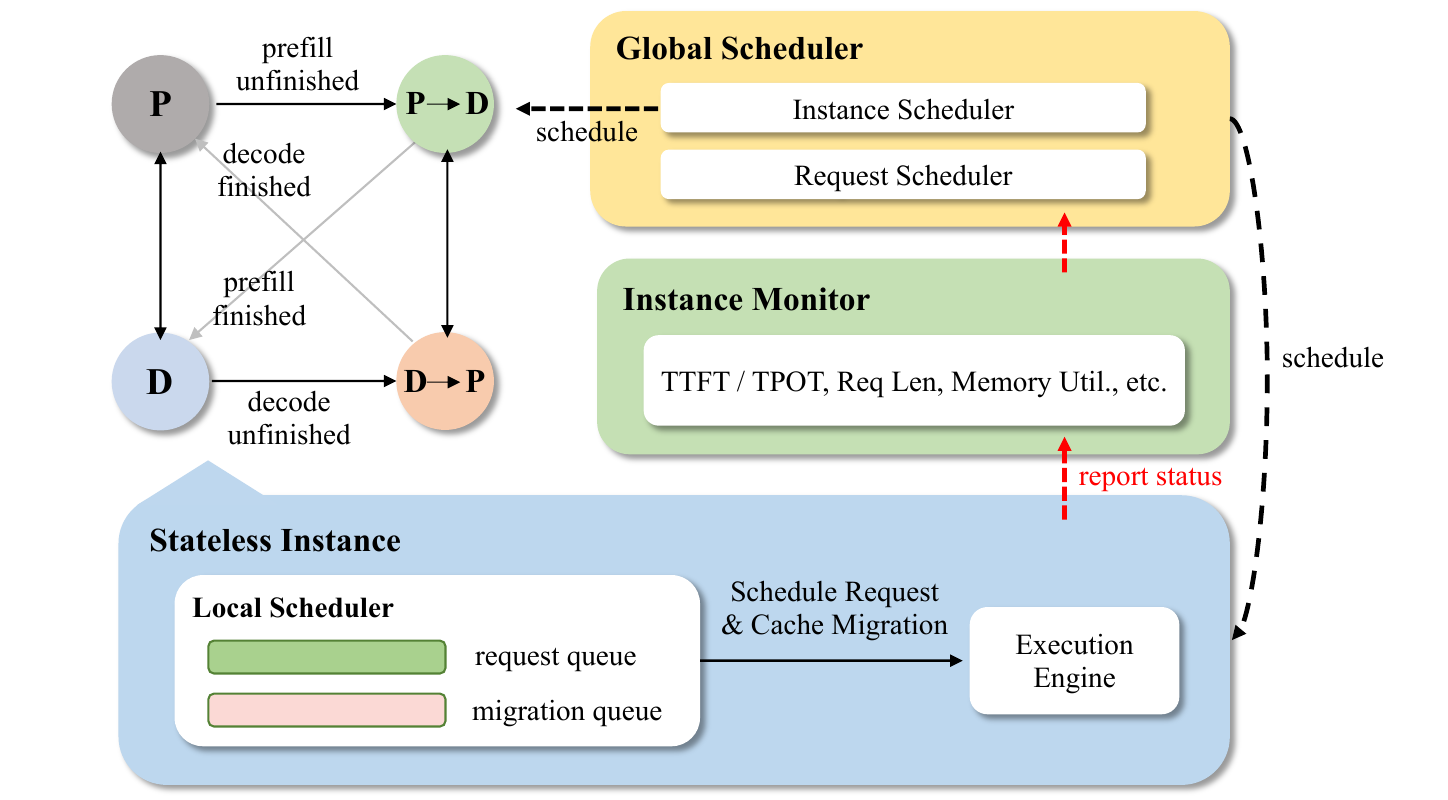} 
  \caption{Overview design of Dynamic PD Disaggregation Scheduler Policy.}
  \label{fig:pd-overview}
\end{figure}

\paragraph{Inefficiency of Existing PD Disaggregation Policy under Workload Fluctuations.} 
The Prefill-Decode (PD) disaggregated inference architecture~\cite{distserve, infer-metric, qin2025mooncake, pd-serve} partitions computing instances into dedicated prefill and decode instances, each handling their respective processing phases. This design mitigates interference between prefill and decode requests, achieving superior performance compared to PD co-located architectures~\cite{distserve}. 

However, we observe that most existing PD disaggregated systems employing static instance partitioning schemes suffer from low hardware utilization and inadequate responsiveness to traffic bursts.
On one hand, existing analysis-based methods~\cite{distserve, infer-metric, qin2025mooncake, pd-serve} typically determine the PD ratio using profiling or simulator data. These approaches remain effective only when request arrival patterns and length distributions remain relatively stable. Under significant workload fluctuations, pre-collected analysis data often fails to accurately capture real-time request characteristics, leading to mismatches between preset load-balancing ratios and actual load requirements. On the other hand, when confronting substantial variations in input/output lengths within production workloads, existing PD disaggregation architectures~\cite{distserve, splitwise} generally adopt dynamic instance type adjustment strategies~\cite{loongserve}. Nevertheless, online switching between prefill and decode instances typically involves multiple steps—including monitoring, waiting for flip conditions, and instance restarting—which introduces substantial latency overhead~\cite{tetriinfer}. To address these limitations, we propose a Dynamic PD Disaggregation Policy that adaptively adjusts resource allocation in response to real-time workload characteristics.

\paragraph{Runtime Instance Monitor.} Performance indicators such as TTFT and TPOT directly reflect the capability of instances to process requests in the prefill and decode phases, respectively. They serve as the core basis for evaluating whether an inference system can meet SLO requirements. Ideally, prediction-based methods could be adopted to dynamically assess these indicators. However, while TTFT exhibits relatively predictable characteristics (as its computation time is proportional to the square of the input sequence length~\cite{ecoserve, pd-serve}), due to the uncertainty in the length of output tokens and transfer overhead, TPOT is difficult to accurately predict using traditional input/output features and cluster load metrics. To address this issue, we deploy additional instance monitors to collect real-time performance data from each computing instance. This data includes metrics such as the number and length of prefill and decode requests, memory usage, TTFT, TPOT, and token generation intervals. The system can further use these real-time performance indicators to dynamically evaluate instance loads and adjust scheduling strategies. 

\paragraph{Stateless Instance and Elastic Pool.} As illustrated in Figure~\ref{fig:unified-scheduler}, Dynamic PD Disaggregation Policy adopts a design of stateless instances and elastic instance pools to enable fast and dynamic switching of instance roles. First, it treats prefill or decode phase as the request attribute rather than instance attribute. Instances are designed to be stateless, allowing each instance to process both prefill and decode requests simultaneously. Additionally, to facilitate the management of multiple instances, we further extend PD pools to four elastic instance pools (i.e., P, D, P$\rightarrow$D, D$\rightarrow$P) as described in \textsection\ref{sec:overview-service}.
When flipping an instance (switching its role), we only needs to remove the instance from its original pool and move it to the new pool. This achieves zero-wait-time instance scheduling, avoiding the overhead of instance restart or model reloading incurred in traditional systems.

\paragraph{SLO-aware Instance Role Switching.} 
The instance scheduling strategy is dynamically adjusted strictly based on SLO objectives: during the prefill phase, if it is predicted that existing instances cannot meet the TTFT requirements, the conversion of decode instances is triggered; while during the decode phase, when resource shortage occurs, the average token generation interval exceeds the TPOT threshold, or prefill instances are idle, the conversion of prefill instances to decode instances will be initiated to cope with sudden traffic surges. Specifically, when decode instances are reallocated to prefill instances, the scheduler prioritizes selecting the instance with the lightest load (i.e., the fewest tokens being processed) from the P$\rightarrow$D pool for role conversion, and always ensures that at least two decode instances are available; conversely, when prefill instances are reallocated to decode instances, the scheduler prioritizes scheduling from the D$\rightarrow$P pool, which avoids local overload and maximizes resource utilization.

\paragraph{SLO-aware Request Scheduling.}
The request scheduling scheme adopts a two-level architecture: 

$\triangleright$ \, Global Request Scheduler: The scheduler implements a greedy strategy of prioritizing the lightest load, while being strictly restricted by SLO constraints. For prefill requests, the scheduler first evaluates the estimated queuing latency of each instance in the Prefill pool, selects the candidate instance with the smallest latency, and then invokes the TTFT prediction model for verification: if the estimated TTFT can still meet the SLO requirements after assigning the request to this instance, the request is allocated immediately; otherwise, it continues to find a suitable instance in the D$\rightarrow$P pool. If no instance can meet the TTFT requirements, the instance scheduling mechanism is triggered to allocate resources from the decode side. For decode requests, the scheduler gives priority to having the original prefill instance continue processing (to avoid KV Cache transfer overhead); secondly, it selects the instance with the fewest running tokens in the Decode pool, and checks whether the total number of tokens in its current batch is below the memory capacity upper limit and computing throughput limit determined by pre-analysis, so as to ensure that the new request will not cause TPOT to exceed the standard.

$\triangleright$ \, Local Request Scheduler: Each instance adopts a refined queue management strategy internally. KV Cache transfer events are placed in an independent migration queue and processed sequentially in accordance with the FCFS principle; for forward requests, an innovative scheme combining Chunked Prefill~\cite{chunk-prefill} and Continuous Batching~\cite{vllm} is adopted: on the premise of ensuring that decode requests are prioritized to enter the running batch, the remaining computing resources are used to process chunked prefill requests in parallel~\cite{vllm-v1-schedule}.

\subsection{Hybrid EPD Disaggregation Scheduler Policy}
\label{sec:epd-planner}

\paragraph{Challenges of Multi-modal Inference.}
The inference process of multimodal large language models (MLLMs)~\cite{qwen-vl, qwen2.5-vl, minicpm, deepseek-vl, mllm-survey, llava} typically consists of three phases: the image encoding phase (for extracting image features), the prefill phase (for encoding images and text prompts, feeding them into the language model to generate the first output token, and caching intermediate states), and the decode phase (for iteratively generating subsequent tokens based on the cached data). Existing mainstream inference engines, such as vLLM~\cite{vllm}, Text-Generation-Inference~\cite{tgi}, SGLang~\cite{sglang}, and DistServe~\cite{distserve} are all tailored for LLMs, and thus face several challenges in handling inference tasks for MLLMs:

$\triangleright$ \, Insufficient Parallelism: For example, the inference of visual models and language models can be executed in parallel, thus improving the utilization of computing resources. However, most existing inference engines \cite{ning2024inf} adopt a serial strategy, failing to exploit the inter-request parallelism effectively. 

$\triangleright$ \, Coarse-grained Scheduling: The decode phase is a memory-intensive task, which is suitable for batch processing to improve throughput~\cite{d-memory}; the prefill phase is computationally intensive, and it is appropriate to adopt the chunked prefill \cite{chunk-prefill} together with the decode phase to balance latency and throughput~\cite{p-compute}. The computation and memory access overhead of the encode phase lies between the two, and it can also benefit from independent scheduling and batch processing. Nevertheless, existing engines process encode and prefill in a combined manner, failing to perform phase-specific batch processing and lacking support for chunked prefill. Coarse-grained scheduling makes it difficult to finely control the execution time. 

$\triangleright$ \, Disaggregation Strategy: Existing architectures such as DistServe~\cite{distserve} reduce resource interference by decoupling the prefill and decode phases. However, in multimodal inference, encoding and prefill jointly affect TTFT, while prefill and decode jointly determine TPOT. Under different loads, how to select the optimal decoupling strategy remains a challenge~\cite{singh2024efficiently}. For example, the performance of strategies such as E+P+D, EP+D, and ED+P in different tasks has not yet been evaluated, and in-depth analysis is required to guide system design.

\begin{figure}[t]
  \centering
  \includegraphics[width=0.95\textwidth]{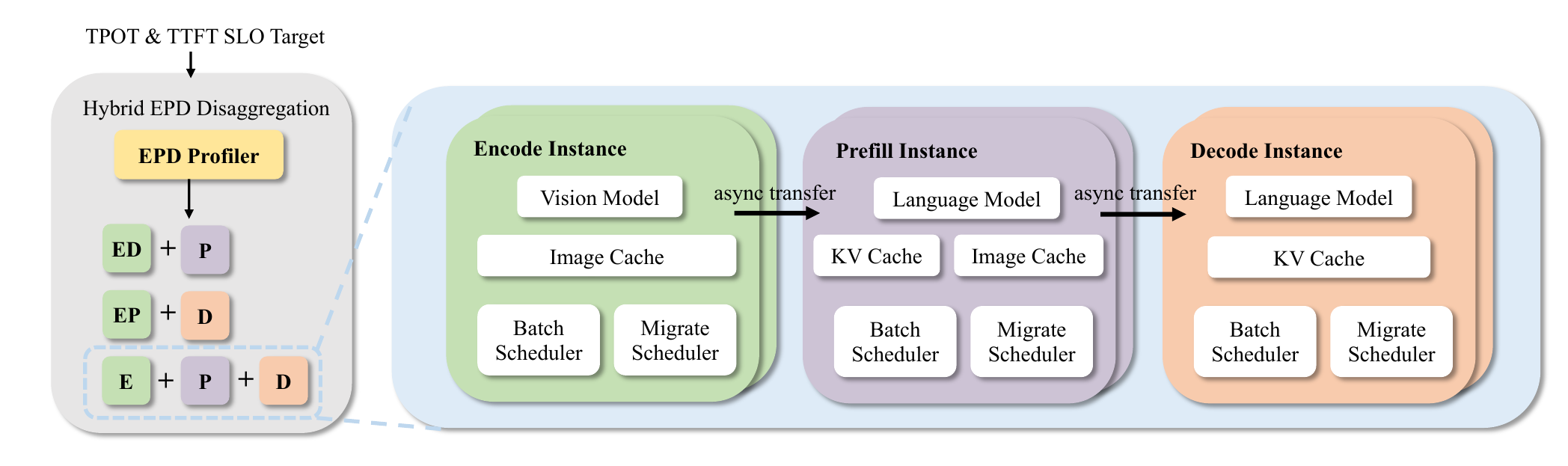} 
  \caption{Overview design of Hybrid EPD Disaggregation Policy.}
  \label{fig:epd-planner}
\end{figure}

\paragraph{Dual-stream Parallel.} As illustrated in Figure~\ref{fig:epd-planner}, we adopt a dual-stream parallel mode, where the visual model and language model are assigned to separate execution streams: the visual stream is dedicated to performing computationally intensive image encoding tasks, while the language stream handles the prefill and decode operations for language generation. By isolating workloads into distinct streams, our system achieves concurrent execution of heterogeneous phases from different requests.

\paragraph{Three-Phase Disaggregation.} To address the differences in computation and memory access characteristics among tasks in each phase of multimodal inference, we propose a phase-aware scheduling strategy. Requests are subdivided into three phases within an instance: encode, prefill, and decode. Batch processing and scheduling optimizations are performed for each phase respectively: 

$\triangleright$ \, Optimized Batch Processing: The maximum batch size for image encoding and the token budget for the language model are set according to the user's SLO. Specifically, during system startup, we use a binary search method to profile the maximum encoding batch size and the model's token budget, ensuring that the execution time of subsequent batch processing tasks in each iteration is less than the TPOT SLO. In each iteration, we (i) first add all running decode requests to the current batch; (ii) then check for any chunked prefill tasks that have been partially computed. If such tasks exist, we add them to the batch; (iii) if there are no chunked prefill tasks, we check for pending encoding tasks and add them if any exist. This approach aims to complete requests in the prefill phase as quickly as possible, reducing their TTFT. New requests' encoding phases are processed only when no requests are in the prefill phase.

$\triangleright$ \, Phase-aware Scheduling: To address the difficulty in selecting decoupling strategies for multimodal inference tasks, xLLM proposes the Hybrid Encode-Prefill-Decode Disaggregation architecture, an innovative multi-instance collaborative inference framework. In this architecture, each instance only executes part of the subtasks among the three phases, while the remaining phases are processed by migrating requests to other instances, thereby avoiding resource waste and interference. During runtime, our system automatically selects the optimal disaggregation strategy based on historical request profiling (via \textit{EPD Profiler}) to dynamically balance throughput and latency objectives. Compared with the traditional approach of binding instances to single-phase or full-phase tasks, this architecture significantly improves the overall system processing capability and resource utilization efficiency while meeting SLO requirements.

\subsection{Global KV Cache Management}
\label{sec:cache-mgr}

During the decode phase of LLMs, subsequent tokens are generated autoregressively one by one. Although the computational cost per step is relatively low, frequent access to the historical KV Cache is required, making memory bandwidth the primary bottleneck. As model size expands and context window grows, the memory consumption of the KV Cache exhibits an exponentially increasing trend. For instance, a context of 128K tokens may consume over 40GB of memory, severely straining the memory resources of single-GPU devices~\cite{deepseek}.

\begin{figure}[t]
  \centering
  \includegraphics[width=0.8\textwidth]{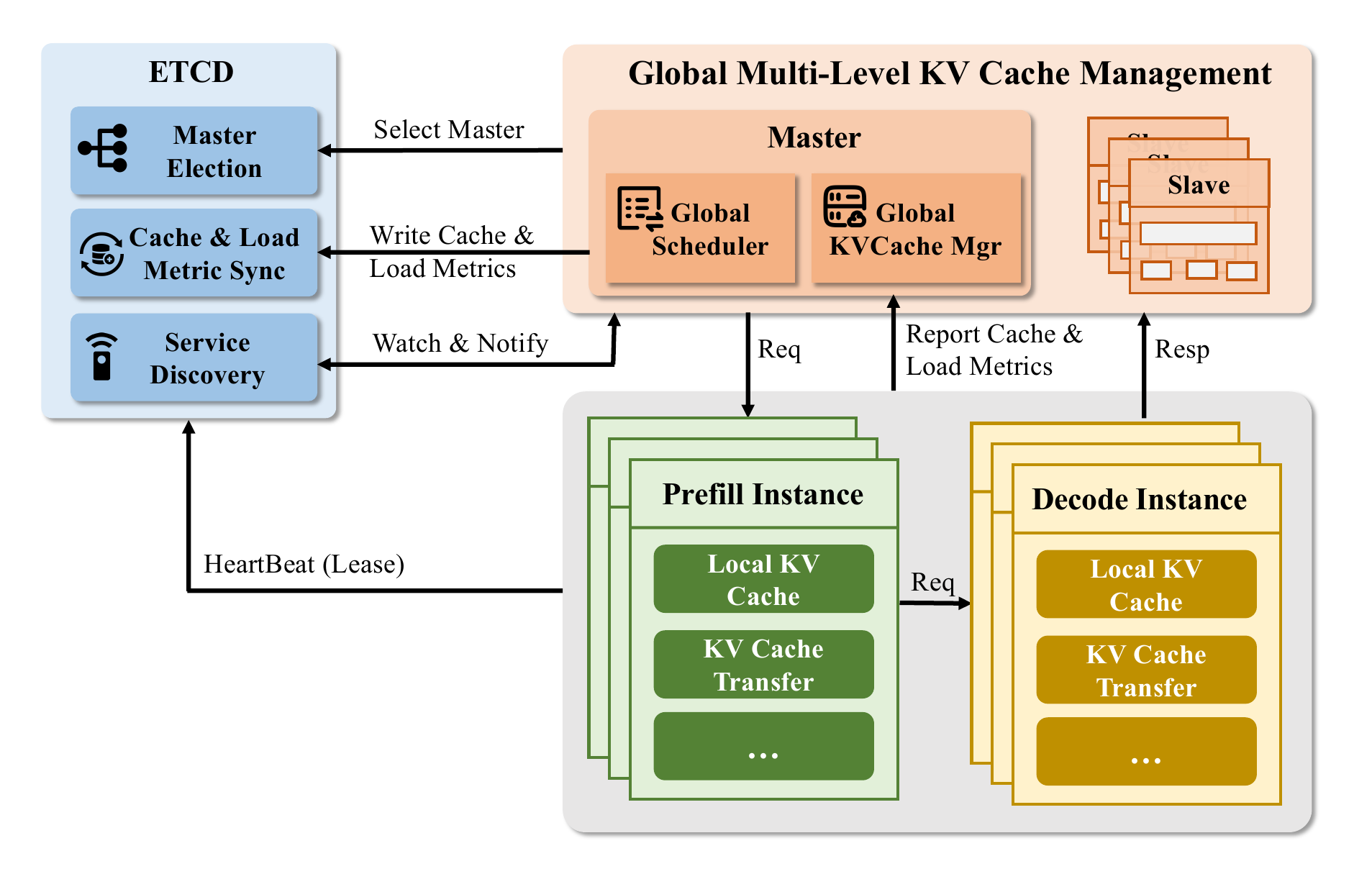} 
  \caption{The framework of Global Multi-Level KV Cache Management.}
  \label{fig:global_kv_cache}
\end{figure}

Although current mainstream optimization solutions like vLLM~\cite{vllm} and FasterTransformer~\cite{fastertransformer2023} have made significant progress in single-instance environments, many critical issues remain unresolved. In long-context scenarios, the time required for the prefill phase increases dramatically, while the decode phase suffers from intense competition for memory bandwidth. To meet stringent SLO requirements (e.g., TTFT $< 2$s, TBT $< 100$ms), one instance often has to reserve excessive resources, instead of leveraging resources in other instances. This significantly restricts the overall cluster resource utilization.
To address these challenges, we propose a global multi-level KV Cache management system with a memory-compute integrated architecture, as illustrated in Figure~\ref{fig:global_kv_cache}.

\paragraph{Distributed Storage and KV Transfer.} 
We adopt the Mooncake Store~\cite{qin2025mooncake}, a KV cache-centric storage engine, as the underlying storage infrastructure for xLLM's KV Cache, along with the Mooncake Transfer Engine as the transmission component. Mooncake Store leverages striping and parallel I/O techniques to fully utilize the aggregated bandwidth of multiple network cards. It provides three persistence strategies---Eager, Lazy, and None---to meet data durability requirements across various scenarios. Additionally, Mooncake Store offers flexible object storage capabilities, supporting multiple replicas and eventual consistency, which effectively mitigates hot-spot access pressure. 
As the core transmission engine of the system, Mooncake Transfer Engine~\cite{qin2025mooncake} automatically selects the optimal transmission path based on data location and abstracts low-level complexities through a unified Segment and BatchTransfer interface.

\paragraph{Multi-Level KV Cache Management.} 
At the global level, the system employs ETCD~\cite{etcd} as a metadata service middleware to achieve cluster service registration, load information synchronization, and global cache state management. Each computing instance maintains a local multi-level KV Cache pool (HBM $>$ DRAM $>$ SSD) adhering to the strict consistency rule: ``if data resides in HBM, it must also be present in DRAM''. Specifically, when local operations involving KV cache (including prefix cache~\cite{vllm}) loading and offloading occur, these operational events are aggregated at regular intervals and transmitted to the xLLM-Service via ETCD heartbeat mechanisms, enabling unified global monitoring and management.

\paragraph{KV Cache-aware Scheduling.} In terms of scheduling strategy, the system implements a decision-making mechanism based on KV cache utilization, which operates through three key steps: (1) Prefix Matching Detection: calculating the KV cache reuse rate for each candidate node via prefix matching analysis; (2) Performance Estimation: estimating the expected latency for different nodes according to current load conditions and cache hit rates; (3) Optimal Node Selection: identifying the node with the optimal overall performance for request processing, thereby enabling dynamic offloading and migration of KV cache.


\subsection{Fast Fault Recovery Architecture}
\label{sec:fault-tolerance}
Current fault handling mechanisms are mainly designed for model training \cite{failover-train, flashrecovery, wu2023transom} and cannot be directly applied to large model inference due to its low-latency requirements. Specifically, existing approaches primarily adopt a checkpoint-then-recovery method~\cite{duan2024efficient, mohan2021checkfreq, chen2023cost} for fault handling, which periodically stores model data as checkpoints in distributed storage and reloads the most recent checkpoint after a failure occurs~\cite{hw-recovery}. As model parameters increase, the overhead of storage and loading gradually grows. This is still acceptable in training since there are no strict latency requirements~\cite{bytecheckpoint}, but in inference, it may cause all high-priority requests on the failed instance to time out, resulting in severe losses.

To address this issue, we propose an efficient failover architecture specifically for large model inference, with targeted optimizations in two aspects: fast request migration and fast instance recovery. Fast request migration ensures the SLO of high-priority tasks mainly through an efficient kv cache quick recovery strategy, while fast instance recovery achieves low-overhead recovery through efficient masking of computation and communication. Through the above optimization methods, we achieve fast fault recovery, greatly reducing the performance and economic losses caused by faults.
\section{xLLM-Engine Designs}
\label{sec:engine}

\subsection{Multi-layer Pipeline Execution Engine}
\label{sec:pipeline}
\begin{figure}[t]
  \centering
  \includegraphics[width=1.0\textwidth]{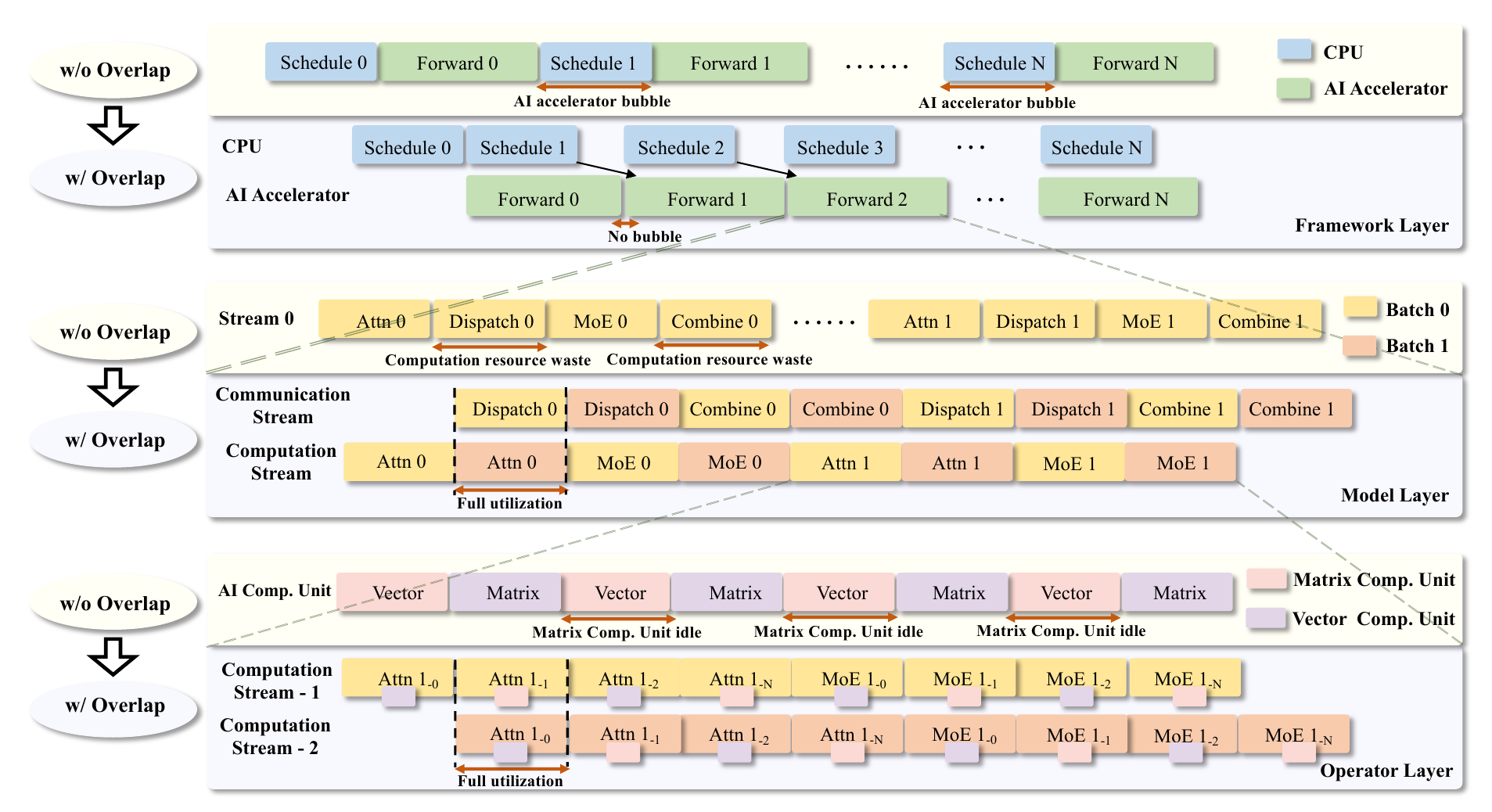} 
  \caption{Multi-layer pipeline execution engine.}
  \label{fig:pipeline}
\end{figure}

Traditional autoregressive LLM inference relies on single-stream sequential execution. This conventional approach fails to fully exploit the parallelism capabilities of modern hardware, often leading to computational resource underutilization, communication stalls, and cascading blocking delays in heterogeneous computing environments~\cite{nanoflow, tokenweave}.
As illustrated in Figure~\ref{fig:pipeline}, the system's performance is hindered by three primary sources of inefficiency:
(1) \textit{CPU-Accelerator Dependency:} At the framework level, a rigid dependency between the CPU and the accelerator forces the accelerator to remain idle during task scheduling and data processing phases, creating significant ``computation bubbles''~\cite{nanoflow, deepseek}.
(2) \textit{Communication Latency:} In both distributed workloads and complex model layers, communication delays interrupt continuous computation, preventing the full utilization of available hardware resources~\cite{tokenweave, iso}.
(3) \textit{Architectural Bottlenecks:} On certain AI accelerators, the compute-focused (tensor cores) and general-purpose (vector cores) computation units lack a shared high-level cache (e.g., L1 or SRAM). This architectural separation necessitates additional data transfers between them, which introduces latency and leads to the underutilization of the specialized computation units.

To address these challenges, we propose a three-tier asynchronous pipeline parallelism design spanning the framework scheduling layer, the model graph layer, and the operator layer, as shown in Figure~\ref{fig:pipeline}, with the goal of maximizing hardware efficiency. 

\paragraph{Framework-Layer Scheduling-Execution Overlap.}
In a typical sequential execution procedure, the accelerator remains idle while the CPU performs scheduling to prepare the input data batch for the next computation cycle. This serial dependency creates significant latency and underutilizes the accelerator.

To eliminate this bottleneck, we introduce an asynchronous pipelined scheduling mechanism that decouples CPU scheduling from accelerator execution. The core idea is to overlap these two stages: while the accelerator is executing the forward pass for the current batch, the CPU concurrently assembles the batch and prepares the metadata for the next batch. 
This parallel workflow effectively hides the latency of CPU-side scheduling and data preparation. The process is as follows:
(1) Accelerator Compute: The accelerator is busy executing the forward pass for the current iteration, which will eventually produce output. (2) CPU Schedule: Instead of waiting, the CPU immediately begins scheduling the batch for the next iteration. To do so, it uses a set of placeholder tokens that stand in for the yet-to-be-computed output. This allows all CPU-side batch preparations to occur concurrently with the accelerator's computation. (3) Seamless Transition: Once the accelerator completes the forward pass and the real output is generated, a fast replacement operation swaps the placeholder tokens with the real generated tokens. Since all scheduling is already complete, the accelerator can begin the next forward pass with minimal delay.

By transforming the conventional serial ``prepare-then-compute"" workflow into a parallelized pipeline, our method effectively masks scheduling latency behind accelerator computation time. This design eliminates idle execution ``bubbles"" in the accelerator timeline, thereby maximizing hardware utilization and enhancing overall system throughput.

\paragraph{Model-Layer Computation-Communication Overlap.}
Maximizing the overlap between computation and communication is a critical design principle in recent LLM frameworks \cite{deepseek,lina}. This strategy is essential for hiding the significant latency of data transfers and achieving high hardware utilization. However, architectural constraints can make this challenging. For instance, on a typical GPU, compute-focused units (Tensor Cores) are tightly coupled with general-purpose units (CUDA Cores) within a Streaming Multiprocessor (SM). Consequently, when an SM is allocated to a communication-related task, its powerful Tensor Cores often sit idle, leading to wasted compute potential.
In contrast, our target accelerator architecture allows for more flexible, independent allocation of its Cube (matrix) and Vector (general-purpose) units. We leverage this flexibility to address the aforementioned challenges by introducing a dual-stream pipeline architecture that uses micro-batch splitting to effectively overlap computation and communication.

Specifically, our architecture consists of a Computation Stream for compute-bound tasks (Attention, ExpertForward, \textit{etc.}) and a Communication Stream for data distribution and collection tasks (MoE Dispatch and Combine). 
To enable pipelining and maximize overall pipeline throughput, we partition a macro-batch $B$ into $n$ micro-batches~\cite{nanoflow} $\left\{ b_1, b_2, \cdots, b_n \right\}$. The two streams asynchronously execute tasks for different micro-batches. 
On the one hand, our scheduling policy dynamically determines the optimal execution order of the micro-batches.
On the other hand, our resource scheduling policy adaptively allocates an appropriate number of AI Cores (Vector Cores and Cube Cores) to the Communication Stream and Computation Stream respectively. 
Figure~\ref{fig:pipeline} illustrates the case of $n=2$, where the Communication Stream performs the Dispatch operation for micro-batch $b_k$ while the Computation Stream executes the ExpertForward pass for the preceding micro-batch $b_{k - 1}$.

\paragraph{Operator-Layer Matrix-Vector Units Overlap.}
On heterogeneous AI accelerators, the serial scheduling of matrix and vector computation units often results in significant underutilization, as one class of units remains idle while the other is active. 
This inefficiency motivates a strategy of operator-level matrix-vector overlap to fully exploit the hardware's parallel capabilities. However, a naive implementation via coarse-grained parallel scheduling, without a systemic resource coordination mechanism, is also problematic. This approach frequently leads to disordered contention for limited compute units, causing resource fragmentation and access conflicts that ultimately degrade overall performance.

To address these challenges, we propose a dynamic resource allocation mechanism for computation units (Cube and Vector Units) based on real-time load monitoring. This mechanism enables a deep pipeline across heterogeneous compute units by dynamically and adaptively assigning the precise type and quantity of resources required for each concurrent operator. By doing so, it mitigates resource contention and ensures that parallel operators execute within highly overlapping time windows, achieving precise computational overlap and maximizing hardware utilization.

We formulate the dynamic resource allocation as an optimization problem. Let $\mathcal{C}$ and $\mathcal{V}$ be the sets of matrix and vector operators, respectively. Let $x_i$ be the number of matrix units (Cube) allocated to the $i_{th}$ matrix operator ($i \in \mathcal{C}$), and $y_j$ be the number of vector units (Vector) allocated to the $j_{th}$ vector operator ($j \in \mathcal{V}$). For simplicity, we assume all operators begin execution simultaneously, without considering data dependencies or communication latency. 
Based on the known computational workload of each operator, our mechanism seeks to find an optimal resource allocation that minimizes the maximum difference in execution times between any two operators. This objective, which we term the alignment loss ($\mathcal{L}_{align}$), effectively synchronizes the completion time of all parallel kernels. The optimization problem is defined as:

\begin{equation}
\begin{aligned}
\text{argmin}_{x_{i},y_{j}} \mathcal{L}_{align} &= \max_{i \in \mathcal{C},\, j \in \mathcal{V}} \left| T_{i} - T_{j} \right|, \\
T_{i} = \frac{W_i}{\gamma_{\text{Cube}} \cdot x_{i}},\quad T_{j} = \frac{W_j}{\gamma_{\text{Vector}} \cdot y_{j}}, &\quad
\sum_{i \in \mathcal{C}} x_i \leq N_{\text{Cube}}, \quad \sum_{j \in \mathcal{V}} y_j \leq N_{\text{Vector}},
\end{aligned}
\end{equation}

where $T$ is the operator execution time, $W$ is its computational workload, $\gamma_{cube}$ and $\gamma_{vector}$ are the peak performance per unit, and $N_{cube}$ and $N_{vector}$ are the total available matrix and vector units.
The meaning of this optimization objective is to minimize the maximum difference in execution time between any pair of concurrent matrix and vector operators. 

\subsection{Adaptive Graph Mode}
\label{sec:acl-graph}
The performance of LLM inference deployment is often impeded by Host-side CPU overhead, particularly when the computation graph is composed of many fine-grained operators \cite{pytorch-cudagraph}. 
This bottleneck manifests as both significant CPU-Accelerator synchronization latency, measured at $5\sim50 \mu s$ per invocation from frequent kernel launches, and suboptimal hardware utilization due to idle accelerator cycles between these intermittent operator executions. 

\begin{figure}[t]
  \centering
  \includegraphics[width=0.95\textwidth]{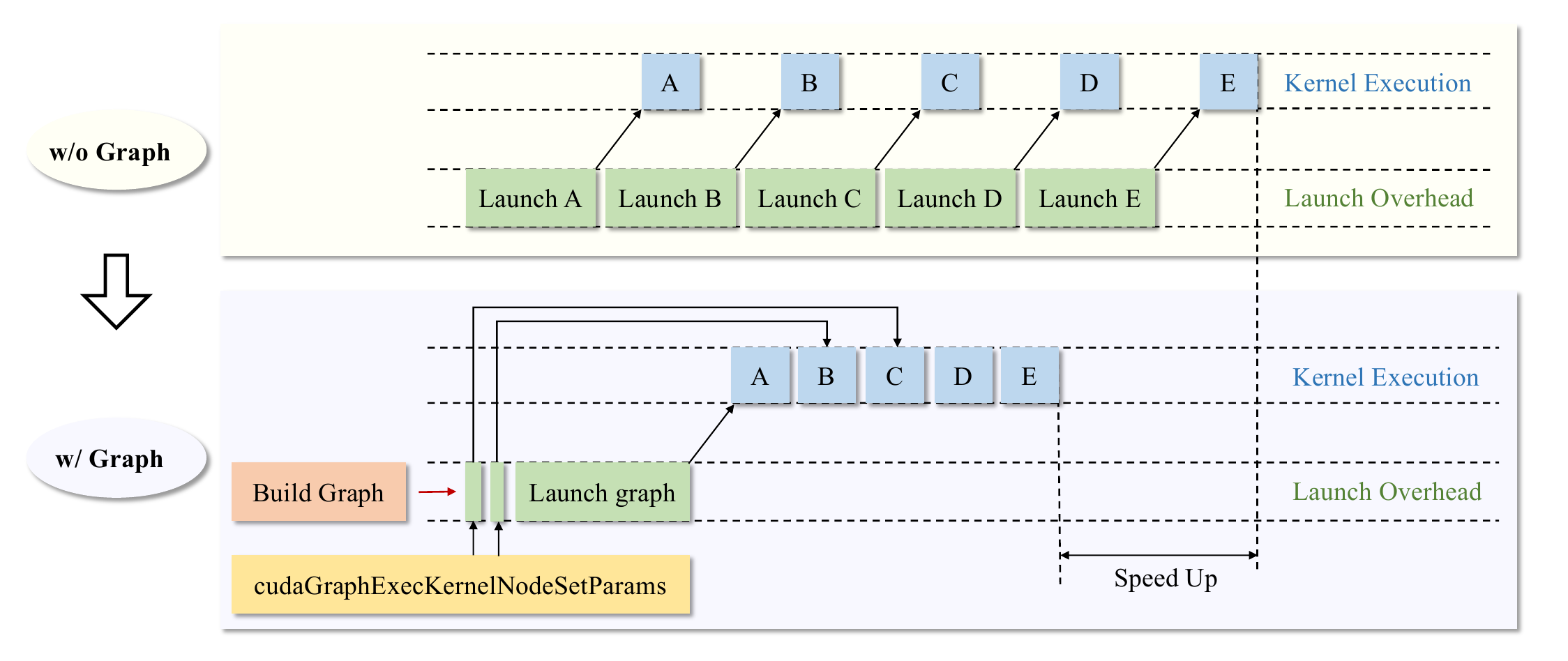} 
  \caption{Pipeline comparison between \textit{Eager Mode} and \textit{Graph Mode}. }
  \label{fig:aclgraph}
\end{figure}

Mainstream AI accelerators such as NVIDIA GPUs (with CUDAGraph)~\cite{nvidia-cudagraph} and Ascend NPUs (with ACLGraph)~\cite{ascend-docs} employ computation graphs to try to handle above issues and enhance host-side scheduling performance. As illustrated in Figure~\ref{fig:aclgraph}, traditional \textit{Eager Mode} relies on the CPU to submit a multitude of small, intensive tasks, leading to frequent launches of small kernels on the accelerator. 
In contrast, the \textit{Graph Mode} enables the CPU to submit one large task, after which the accelerator internally executes the small kernels in a streamlined fashion. This method significantly reduces both launch overhead and ``accelerator bubbles'' (idle cycles). Specifically, the \textit{Graph Mode}, such as ACLGraph~\cite{ascend-docs}, comprises two distinct phases: graph capture and graph execution. During the graph capture phase, the entire computation flow is recorded to capture the sequence of kernel calls and their dependencies, including kernel launch parameters. 
This recorded workflow is then pre-compiled into a replayable directed acyclic graph (DAG) object. It is important to note that during this phase, tasks are merely staged within the model's runtime instance and are not actually executed. In the subsequent graph execution phase, the entire graph is launched via a single CPU call. The accelerator then follows the pre-defined process autonomously, without real-time intervention from the CPU. This effectively consolidates many individual kernel launches into one single graph launch.

Building upon this foundation, we further advance the practical implementation of the ACLGraph~\cite{ascend-docs} and propose our \textit{Adaptive Graph Mode} through addressing key challenges including dynamic shape adaptation, memory reuse conflicts across multiple graphs, and compatibility with custom operators. This enables the pre-compilation of kernel sequences, memory copy operations, and synchronization points into a single computational graph, which is then dispatched to the AI accelerator in one execution. Consequently, our approach significantly reduces kernel launch overhead and minimizes accelerator idle periods, maximizing overall hardware efficiency.

\paragraph{Dynamic Shape Adaptation via Partial Graph and Dimension Parameterization.}

ACLGraph~\cite{ascend-docs} fixes kernel parameters and memory sizes during the graph capturing phase. However, in real-world scenarios, LLMs typically process inputs with variable sequence lengths and batch sizes, making this static capturing characteristic difficult to adapt to dynamic input requirements. To address this issue, dimension parameterization can be employed, treating key dynamic dimensions—such as batch size and sequence length—as input parameters to the entire graph, thereby enhancing flexibility. During memory allocation and kernel configuration, these dynamic parameters are used to compute actual required values; for example, the memory size can be calculated using the formula: $alloc\_size=batch\_size \times seq\_len \times hidden\_size$.
At the graph launch stage, the actual batch size and sequence length are passed as parameters to ensure the graph adapts to varying input dimensions.

Although parameterization can handle variations in dynamic dimensions, certain operator implementations on different hardware may not support dynamic dimension parameterization. To overcome this, we propose the \textit{Partial Graph Mode}: the model is partitioned into modules with simple dynamic shapes (such as FFN and LayerNorm, which only have the $num\_tokens$ dimension as dynamic) and modules with complex dynamic shapes (such as Multi-Head Attention, which involves multiple dynamic dimensions like $batch\_size$, $kv\_seq\_len$, and $q\_seq\_len$). Modules with simple dynamic shapes are extracted and compiled for execution using \textit{Partial Graph Mode}, while those with complex dynamic shapes are executed in \textit{Eager mode}. Our \textit{Adaptive Graph Mode} dynamically selects the most appropriate mode above based on the current input shapes. This intelligent selection ensures optimal performance across varying workload conditions.

Table~\ref{table:dynamic_shape} compares the key characteristics of three dynamic shape processing mode. The traditional \textit{Eager Mode} requires no pre-compilation but incurs high scheduling overhead at runtime due to N kernel launches (where $N$ is the number of operators). The \textit{Full Graph Mode}, by fixing shapes, allows single compilation and very low launch overhead, yet its lack of dynamic adaptability limits its applicability. 
In contrast, the proposed parameterized and multi-caching Partial Graph Mode strikes an optimal balance. It achieves execution performance comparable to that of a full graph while maintaining high flexibility, by trading a manageable number of pre-compilations ($M \ll N_{req}$, where $M$ is the number of cached graphs and $N_{req}$ is the number of actual requests) for the high efficiency of a single graph launch.


\begin{table}[t]
\centering
\resizebox{\textwidth}{!}{
\begin{tabular}{c|c|c|c}
\toprule
\textbf{Method} & \textbf{Eager Mode} & \textbf{Full Graph Mode} & \textbf{Partial Graph Mode} \\
\midrule
Compilation Times & $0$ & $1$ & $M$ \\
\midrule
Low Launch Overhead & \bad & \good & \good \\
\midrule
Low Memory Usage & \good & \bad & \good \Large /\bad \\
\midrule
High Flexibility & \good & \bad & \good \\
\bottomrule
\end{tabular}
}
\vspace{2mm}
\caption{Comparison of different shape handling solutions. Our \textit{Adaptive Graph Mode} dynamically selects the most appropriate mode above based on dynamic shapes.}
\label{table:dynamic_shape}
\end{table}

\paragraph{Efficient and Reusable HBM Memory Pool. }
During the graph capture process, the specific virtual addresses of input, output, and intermediate tensors are recorded. To prevent illegal memory access, which can occur when the address of an input tensor changes during an actual inference request, we develop a shared HBM memory pool. 
First, during graph initialization, a sufficiently large, contiguous block of HBM memory is pre-allocated to serve as the graph's memory pool. Subsequently, for internal address management, computation graph intercepts all tensor memory operations and re-describes them using offsets relative to the pool's base address, rather than absolute virtual addresses.

Before the graph is launched, the user-provided input tensor is copied into the HBM memory pool at the offset pre-defined for the graph's input. After graph execution is complete, the data from the output tensor is copied to the user-specified output buffer address. 
Furthermore, internal tensors within the graph are also managed using fixed offsets within the pool, ensuring safe reuse and efficient memory management.
This managed memory pool guarantees that all addresses used internally by the graph are known and safe, which adapts to external address changes with only the overhead of memory copies before and after launch.

\paragraph{Integrating Custom Operators.}
Integrating performance-critical custom operators, such as PageAttention and AllReduce, into the \textit{Adaptive Graph Mode} presents a unique challenge. The internal implementation of these operators often relies on CPU to perform just-in-time shape calculations and kernel configurations based on runtime inputs. This dynamic behavior conflicts with the pre-compiled nature of computation graph.

To resolve this, we modified the implementation of such operators. The first step is to identify custom operators within the graph that have dynamic shape dependencies. We then refactor these operators to accept shape-related parameters (\textit{e.g.}, dimension sizes or loop counts, which can only be determined at runtime) directly as kernel arguments. This approach avoids hard-coding these values on the host side during the graph capture phase. 
As a result of this modification, these custom operators can be successfully captured by our computation graph. By leveraging the graph's parameterization mechanism, the required dynamic arguments for these kernels can be derived from the graph's main input parameters and passed along during execution. This method achieves seamless integration of essential custom operators within the \textit{Adaptive Graph Mode} framework.

\subsection{Efficient Memory Management}
\label{sec:xtensor}

LLM inference deployment faces a critical balance between memory utilization and computational efficiency, particularly in autoregressive generation tasks where efficient KV Cache management has become a key challenge. Traditional solutions fall into two categories: First, contiguous memory allocation approaches that statically pre-allocate memory space based on maximum sequence length before inference, ensuring physical contiguity of KV Cache. While this improves computational efficiency, it results in low memory utilization. Second, PagedAttention~\cite{kwon2023efficient} approaches that support larger batch sizes through paging mechanisms, but frequent access to block tables sacrifices computational efficiency, and the increased parameters complicate operator development and debugging. To address this challenge, inspired by virtual memory management in operating system field \cite{ellm, prism}, we propose the xTensor memory management scheme, which adopts a "logically contiguous, physically discrete" KV Cache storage structure that resolves the contradiction between memory contiguity and dynamic allocation, thereby achieving both high computational efficiency and high memory utilization (as shown in Table~\ref{table:xtensor}).

\begin{figure}[t]
  \centering
  \includegraphics[width=\textwidth]{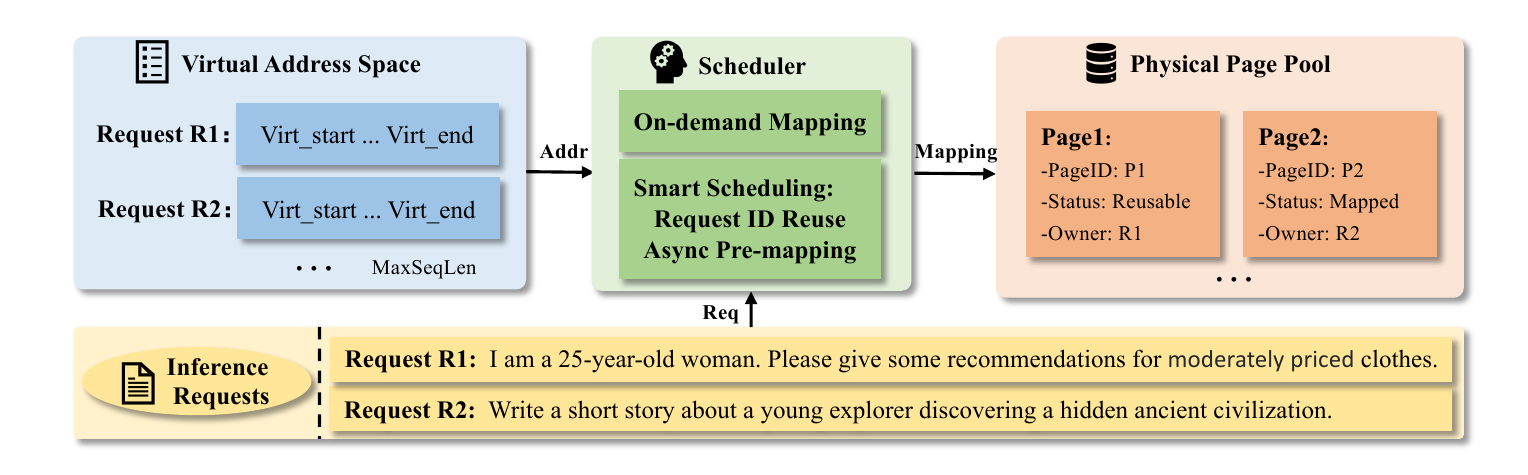} 
  \caption{The framework of xTensor memory management.}
  \label{fig:xtensor}
\end{figure}

\begin{table}[t]
\centering
\resizebox{\textwidth}{!}{
\begin{tabular}{c|c|c|c}
\toprule
\textbf{Features} & \textbf{Contiguous Allocation} & \textbf{Paged Attention} & \textbf{xTensor} (Ours) \\
\midrule
Efficient Memory Usage & \bad & \good & \good \\
\midrule
Efficient Computation & \good & \bad & \good \\
\midrule
Large Batch Support & \bad & \good & \good \\
\midrule
Operator Development Complexity & \good & \bad & \good \\
\bottomrule
\end{tabular}
}
\vspace{2mm}
\caption{Comparison of memory management strategies.}
\label{table:xtensor}
\end{table}

\paragraph{Physical Memory Page Management and Virtual Memory Tensor Management.}
Figure~\ref{fig:xtensor} illustrates the xTensor memory management framework. During service initialization, based on the available memory size for KV Cache, a large number of fixed-size discrete physical memory pages are pre-allocated. Each physical page maintains a triple state $\langle PageID, Status, OwnerSession \rangle$, where $PageID$ represents the page identifier, $OwnerSession$ indicates the service owning the page, and $Status \in \{Free, Allocated, Mapped, Reusable\}$ represents states of free, allocated, mapped, and reusable to dynamically track physical page usage. Subsequently, each inference request is pre-allocated with a logically contiguous virtual address space with an address range equal to $MaxSeqLen$, where this virtual address space is not actually associated with physical pages during allocation. This decoupled design provides operators with a virtual view of KV Cache stored in contiguous memory, thereby masking the discrete nature of underlying physical pages~\cite{vtensor, vattention}.

\paragraph{On-demand Memory Mapping.}
During actual allocation, the scheduler dynamically maps physical memory pages on-demand based on the actual sequence length of requests. When a sequence generates new tokens and KV Cache needs expansion, the scheduler retrieves one or more free pages from the pre-allocated physical page pool and maps these physical pages to the next available contiguous address location in the request's virtual address space. Since physical memory is gradually mapped as sequences grow, short sequences only require a small amount of physical memory, thus avoiding the memory waste of contiguous allocation strategies that still require reserving space according to maximum sequence length for short sequences.

After memory mapping is completed, kernel access to virtual addresses can automatically locate the corresponding physical page $phypage_{idx}$ and $offset$:

\begin{equation}
\begin{aligned}
phypage_{idx} &= \left\lfloor \frac{virt\_addr - virt\_start}{page\_size} \right\rfloor, \\
offset &= (virt\_addr - virt\_start) \bmod page\_size,
\end{aligned}
\end{equation}

where $virt\_addr$ and $virt\_start$ represent the current virtual address and starting virtual address respectively, and $page\_size$ represents the physical page size.

\paragraph{Low Overhead Allocation.}
If virtual address to physical page mapping is immediately released upon request completion, new requests require remapping physical pages, but the $Unmap$ operations on accelerators incur significant overhead. Therefore, under high load conditions, frequent $Map$/$Unmap$ operations become performance bottlenecks, especially for numerous short-term concurrent requests. We adopt request ID reuse and asynchronous pre-mapping designs to reduce latency from $Map$/$Unmap$ operations.

\begin{itemize}[leftmargin=*]
    \item \textbf{Physical Page Reuse.} Upon request completion, physical pages are not immediately released but marked as Reusable. When new requests arrive, if their required KV Cache size matches some Reusable physical page set, that page set is remapped to the new request's virtual address space. This saves expensive $Map$ and $Unmap$ operations through fast virtual address remapping, particularly suitable for scenarios with concentrated request length distributions.
    \item \textbf{Asynchronous Pre-mapping.} During current token decoding, the next token's required physical pages are asynchronously predicted and mapped. Since new token writes to KV Cache occur after current token decoding computation, this design can partially hide physical page mapping operations, significantly reducing page mapping latency.
\end{itemize}

\paragraph{Operator Adaptation.}
The decoupled design of xTensor's physical memory pages and virtual contiguous memory requires the adaptation of existing operators. On one hand, to accommodate KV Cache contiguity in virtual address space, operators no longer require $block\_table$ parameter input. Whether for attention operators or $write\_kvcache$ operators responsible for writing KV Cache during Prefill and Decode phases, only the starting virtual address and related offsets need to be passed in, with the system automatically associating virtual addresses to corresponding physical pages during operation. On the other hand, accelerators lack native contiguous KV FlashMLA~\cite{dao2022flashattention} operators and only support PagedFlashMLA~\cite{dao2022flashattention} operators optimized for paging. Therefore, by reconstructing the PagedMLAttention operator in the accelerator operator library, removing Block Table-related logic such as block table queries and cross-page boundary judgments, and adding functionality for direct computation based on input virtual addresses, we implement FlashMLA operators supporting contiguous virtual address input on accelerators.

\subsection{Algorithm Optimizations}
\label{sec:algo}

\subsubsection{Optimized Speculative Decoding}
\label{sec:mtp}

\begin{figure}[t]
  \centering
  \includegraphics[width=\textwidth]{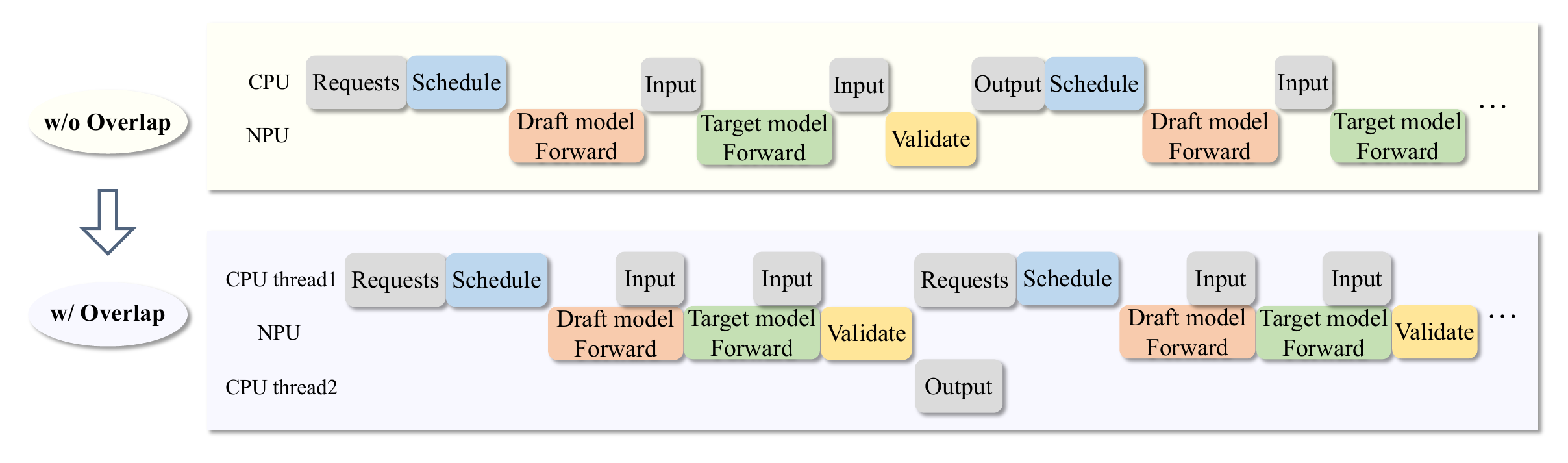} 
  \caption{Comparison between original and our optimized speculative decoding.}
  \label{fig:mtp}
\end{figure}

Traditional autoregressive inference requires LLMs to generate tokens sequentially one by one, resulting in high computational latency and limited throughput. Speculative decoding technology theoretically breaks through this performance bottleneck through a paradigm of parallelly predicting candidate token sequences and quickly verifying them \cite{fastmtp, eagle, mtp}. However, in distributed deployment environments, this technology faces some core challenges:
\textit{First}, synchronous scheduling between CPUs and accelerators in traditional multi-step inference frameworks results in insufficient utilization of computing resources, with accelerators often remaining idle while waiting for data preparation and processing.
\textit{Second}, existing attention mechanisms are not optimized for the characteristics of speculative inference, causing a large amount of redundant data movement.
To address these challenges, xLLM proposes a systematic optimization scheme based on existing frameworks.


\paragraph{Asynchronous Decoding.}
Inspired by~\cite{hw-mtp}, xLLM implements a lightweight asynchronous framework. For timing overlap optimization, while the accelerator executes the main model inference for the current batch, the CPU processes the output decoding of the previous batch and the input preparation for the next batch in parallel—thereby minimizing the idle time of the CPU and accelerator as much as possible.

\paragraph{MLA Optimization.}
xLLM focuses on improving the MLA~\cite{mla,deepseek} computation process. Based on the characteristic that speculative inference needs to process multiple tokens simultaneously on a single sequence, we conducted in-depth optimizations on the self-attention computation in MLA. Specifically, when speculatively inferring m tokens, the self-attention computation of MLA involves product operations between m+1 Q matrices and the same K matrix. By reconstructing the computation process and optimizing the tiling strategy, we effectively reduced the data movement overhead of Q/K matrices. The main optimization measures include:

$\triangleright$ \, Optimization of K matrix loading times: By adjusting the L1 cache allocation scheme to enable parallel loading of multiple Q matrices, and adopting a sliding window-based K matrix loading method that allows consecutive rows of the K matrix to multiply with multiple Q matrices, the number of K matrix loading operations is significantly reduced.

$\triangleright$ \, Q matrix cache residency mechanism: Since there are m+1 Q matrices in speculative inference scenarios, the time to move Q matrices to L1 accounts for a larger proportion of the matrix multiplication process. xLLM redesign the computing scheme to prevent softmax-V product operations from overwriting Q matrices in the L1 cache, enabling Q matrices to remain in the cache and significantly reducing the overhead of repeated data movement, thereby effectively improving the utilization of the tensor core's arithmetic logic units.

\begin{figure}[t]
  \centering
  \includegraphics[width=\textwidth]{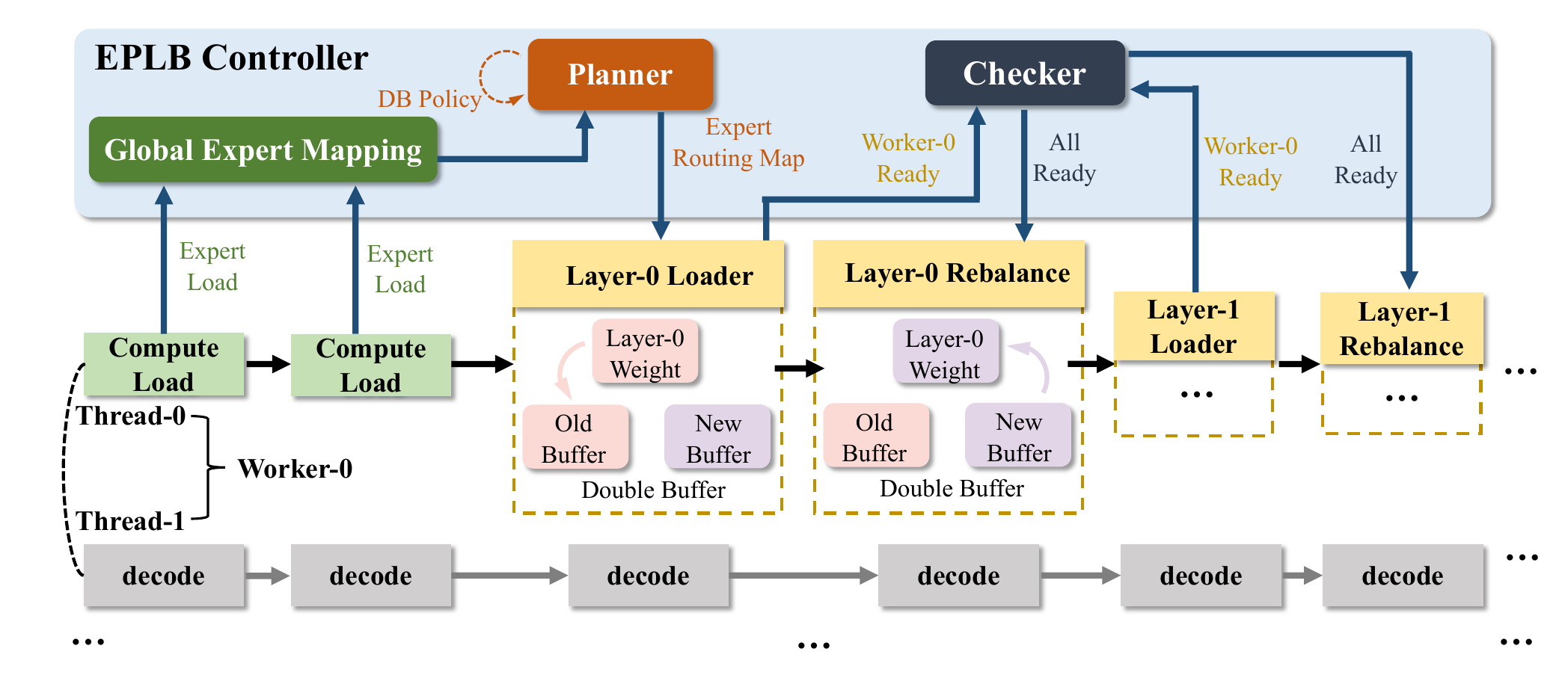} 
  \caption{Overview design of EPLB.}
  \label{fig:eplb}
\end{figure}

\subsubsection{Dynamic EP Load Balance}
\label{sec:eplb}

With the large-scale application of MoE models~\cite{fedus2022switch, dai2024deepseekmoe, qwen3, qwen25} in production environments, their efficient inference capability relies on a expert routing mechanism, which allocates input tokens to different experts for computation~\cite{deepseek, npu, ascend-eplb}. However, in practical deployment, due to the distribution characteristics of input data and the influence of the model's own structure, there may be significant differences in the number of tokens received by different experts~\cite{zeng2025efficientmoe, xue2024moe}. This imbalance leads to low computing resource utilization: some devices are overloaded due to processing too many tokens, while others remain idle, affecting the overall inference efficiency.

To optimize resource utilization, the industry has proposed the Expert Redundancy strategy, which means replicating hot experts and deploying replicas on multiple devices to distribute computing loads. Currently, DeepSeek~\cite{deepseek} adopts two load balancing strategies: group-limited load balance and global load balance, which are optimized respectively for the characteristics of the prefill and decode phases. Adjustments to expert redundancy (such as adding/deleting replicas) require additional memory and may affect inference latency due to weight migration. How to achieve such adjustments efficiently and smoothly is a major challenge. We have made some adaptive improvements based on DeepSeek, realizing dynamic EP load balance.

\paragraph{Expert Load Statistics.}
When the Router distributes tokens, it records the load status of each expert in real time and returns the statistical results through the model output. Each Worker asynchronously starts a thread to periodically aggregate expert loads and report them to the Controller of the Master. The Controller calculates a new routing table based on the load data and then distributes the updated routing table to each Worker.

\paragraph{Double Buffer Mechanism for Weight Update.}
In general, each Worker starts an asynchronous thread to update the weights of single-layer experts during each decoding process. Specifically, after a Device completes the preparation of new weights, it proactively sends a readiness notification to the Controller. The Controller then verifies the weight readiness status of all Worker nodes to ensure global consistency before broadcasting an update instruction, upon which each Worker executes the weight update immediately. Here, we adopt a Double Buffer mechanism, meaning that the old and new expert weights are stored in two separate buffers respectively. The new expert weights are preloaded in the spare memory space, and after the preloading is completed, the address switching is performed to realize the unperceived update of experts.

\subsubsection{Hierarchical DP Load Balance}
\label{sec:dp-balance}

\begin{figure}[t]
  \centering
  \includegraphics[width=\textwidth]{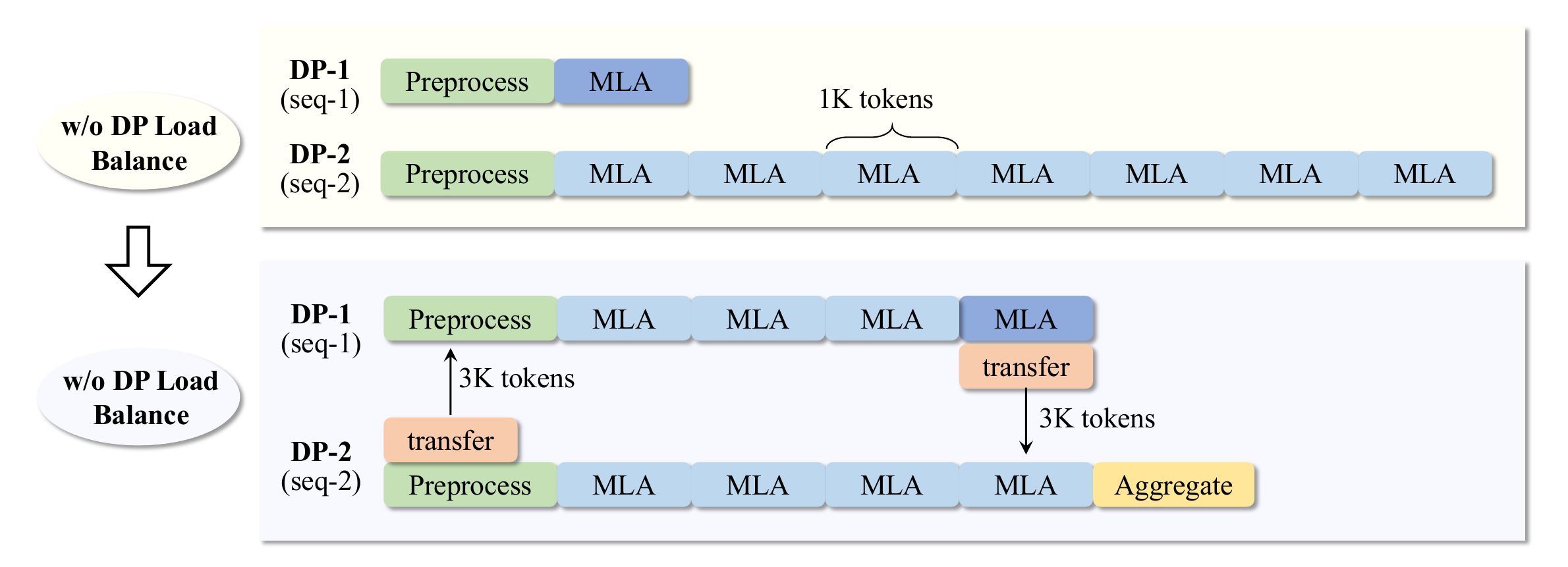} 
  \caption{Inter-DP group load migration at MLA block granularity.}
  \label{fig:dplb}
\end{figure}


In typical MoE model architectures like DeepSeek~\cite{deepseek}, the MoE layers use expert parallelism (EP), while the attention layers use data parallel (DP). This model requires all DP groups to synchronously complete their attention calculations in a single inference step before initiating the MoE all-to-all dispatch operation. This synchronization barrier means the total time for the attention phase is determined by the slowest DP group (the ``straggler''). Faster DP groups cannot immediately proceed to the next stage after finishing their tasks; instead, they are forced into an idle state, waiting for the straggler to complete~\cite{xdeepserve}. This waiting directly translates into wasted computational resources and an increase in overall inference latency.

In practical scenarios, due to the dynamic and unpredictable nature of inference requests, the load within different DP groups is often difficult to balance. Particularly in the decoding phase with large-scale DP (e.g., 80 groups), the load difference between DP groups can reach tens of thousands of tokens, causing fast DP groups to wait for slow ones and wasting approximately 5 milliseconds of delay~\cite{step-3}. Experimental data shows that although the difference in the number of requests processed between DP groups may be as small as four, the corresponding difference in KV Cache token count during the decoding phase can be as high as 20,000. The actual computational workload is directly related to the total number of tokens the system needs to process, as this determines the memory footprint of the KV Cache and the scale of matrix operations in the attention mechanism. This further establishes DP load balancing as a critical factor in determining overall system efficiency.

Current mainstream frameworks like vLLM~\cite{vllm} and SGLang~\cite{sglang} use a simple static round-robin scheduling strategy. Once a request is assigned, its subsequent computations are fixed to a specific DP group, making it unable to adapt to dynamic load changes. Furthermore, at the hardware level, the implementation of operators like the MLA on certain accelerators employs a ``one request per tensor compute core'' partitioning strategy. This can lead to idle compute cores within the same DP group due to differences in request lengths.

To address these challenges, xLLM proposes a hierarchical ``defense-in-depth'' strategy to tackle DP load imbalance issues across different time scales and granularities. The first layer is preventative, KV Cache-aware request scheduling; the second layer performs macroscopic correction through inter-DP group load balancing; and the third layer carries out microscopic correction via intra-DP group load balancing. This layered design enables the system to flexibly address the root causes of various performance degradations, thereby building a robust and efficient inference service.

\paragraph{Layer 1: KV Cache-Aware Scheduling.}
xLLM's first-layer strategy implements request load distribution through KV Cache-aware scheduling. This mechanism moves beyond simple round-robin methods. When a new request arrives, the scheduler comprehensively checks the status of all DP groups, paying special attention to the remaining KV Cache capacity in each. It then preferentially assigns the new request to the group with the most available space. By balancing the total token load and its memory footprint at the system level over the long term, this strategy prevents the formation of severe load imbalances, thereby achieving intelligent resource allocation.

\paragraph{Layer 2: Reactive Inter-DP Group Workload Migration.}
xLLM's second-layer strategy achieves reactive balancing through workload migration between DP groups. During the decoding process, the varying prompt and KV Cache lengths of selected requests cause different computation latencies among DP groups. To counter this, xLLM's central scheduler evaluates the current computational load of each DP group during every inference round. If a significant imbalance is detected, it initiates a workload migration process, moving tasks from overloaded to under-loaded groups. The scheduler also determines the migration granularity -- whether to move an entire batch, a single sequence, or a partial MLA block of a sequence.

Figure~\ref{fig:dplb} illustrates this process at the MLA block granularity. First, xLLM dispatches the input tokens for the migration to both the source and destination DP groups. Then, as all groups execute the MLA Preprocess operation, the request's KV Cache is transferred asynchronously, overlapping communication with computation. Next, all groups begin the attention calculation. The underloaded group prioritizes the migrated attention task, allowing it to immediately send the resulting token back to the source group upon completion while concurrently processing its own native attention operations, further overlapping overhead. Finally, all DP groups use an aggregation operator to gather the external computation results.

\paragraph{Layer 3: Fine-Grained Kernel-Level Optimization.}
xLLM's third-layer strategy achieves fine-grained optimization through kernel-level migration within a DP group. This optimization is performed inside the matrix computation kernels of a single DP group. On one hand, during scheduling, it reorders requests based on their load, replacing the original compute core's round-robin allocation strategy. This aims to keep the total number of computation tokens assigned to each matrix computation unit as consistent as possible. On the other hand, for requests with extremely long sequences, it explicitly splits their computation sequence, migrating parts of the long-sequence request's computation to be calculated with other short-sequence computation units. Through this fine-grained request splitting, it directly resolves the issue of idle compute cores caused by load imbalance within the DP group.

\subsection{Generative Recommendation}
\label{sec:generative}

\begin{figure}[t]
  \centering
  \includegraphics[width=0.9\textwidth]{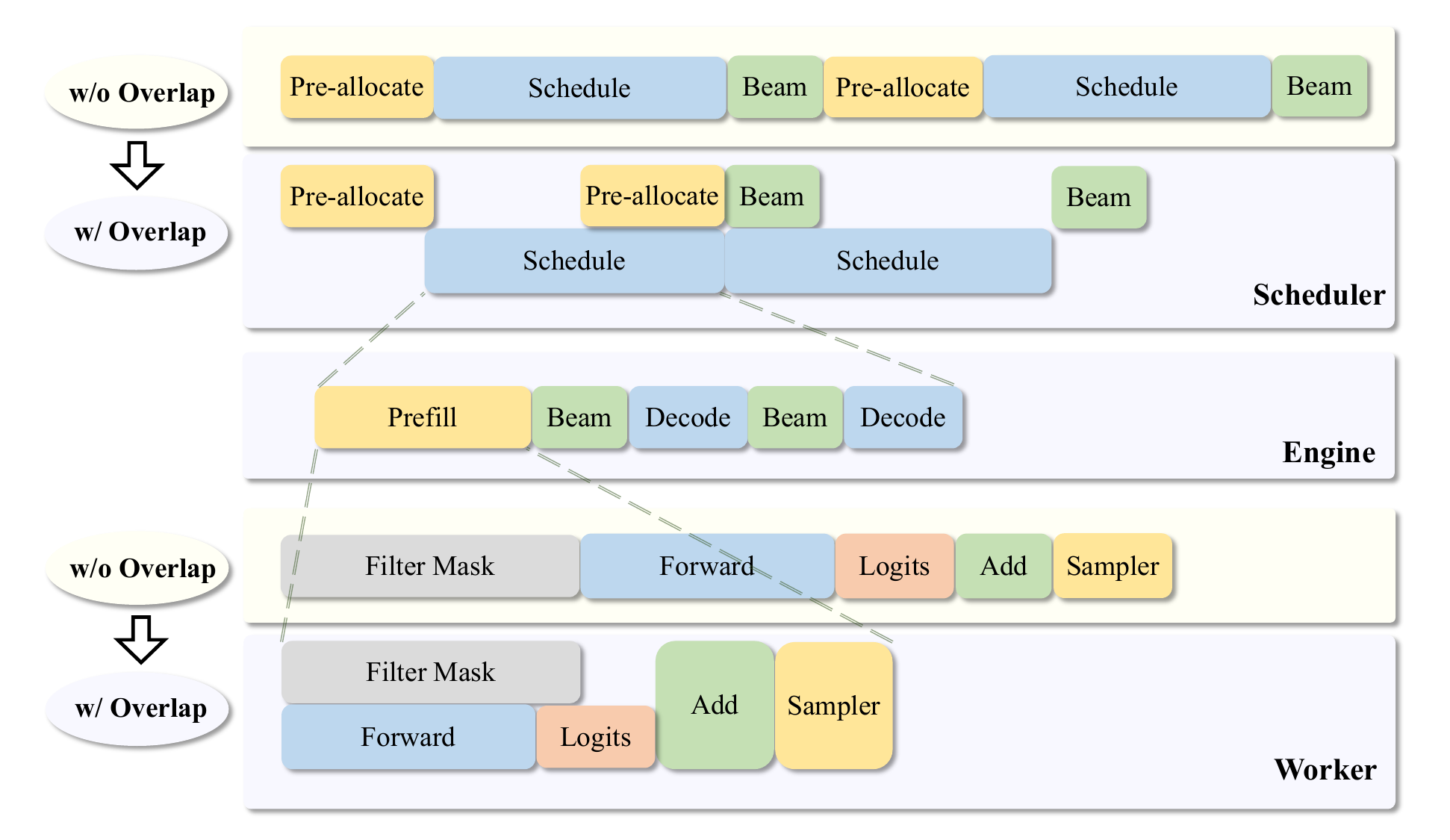} 
  \caption{Overview design of generative recommendation process.}
  \label{fig:generative}
\end{figure}

Recent advancements in generative models have spurred significant interest in generative recommendation \cite{onerec, zhai2024actions, wang2024eager, mtgr, pinrec, hllm}.
Beyond enhancing the retrieval and ranking stages of traditional multi-stage pipelines \cite{hllm,zhai2024actions}, single-stage generative recommendation frameworks directly generate a candidate item list in an auto-regressive manner \cite{onerec}. 
These frameworks usually utilize beam search decoding to directly produce diverse recommendation results, and xLLM implements extensive optimizations for these single-stage frameworks.
As shown in Figure~\ref{fig:generative}, in the generative recommendation scenario, xLLM is dedicated to overlapping host and device operations as much as possible to improve overall execution efficiency. When a request arrives, xLLM preemptively prepares the information required during the forward process on the CPU side. After the preparation is completed, relevant tasks are handed over to the engine for processing.
The Engine executes three forward passes in one go. During this process, the intermediate beam search is a host operation, which by its nature cannot be overlapped with device operations. To optimize the processing pipeline, xLLM internally employs a multi-micro-batch, multi-stream approach.
During the Worker's execution, the CPU operation for generating the filter mask is overlapped with the device operation for logits computation. Through an addition operation, a masked logits is generated before the sampler, thereby achieving data filtering and improving the accuracy and effectiveness of the recommendation results.

\subsubsection{Host-side Optimization}

In scenarios with large $beam\_width$ and $top\_k$ parameters, the significant increase in the number of candidate sequences results in substantial sorting and filtering operations on the host side. This shifts the computational bottleneck of the beam search algorithm from the AI accelerator to the CPU, ultimately leading to a severe CPU-bound issue. The following outlines xLLM's efforts to optimize the beam search process in generative recommendation scenario.

\paragraph{Beam Search Optimization.}
In many fields such as natural language processing, beam search optimization is an efficient algorithm widely used in sequence generation tasks \cite{tree-attention}. When processing a large number of candidate sequences, one of the core operations of beam search is to select $beam\_width$ sequences from numerous candidates. This process is essentially a partial sorting operation, with the unique characteristic that it only needs to identify the top $beam\_width$ elements without fully sorting all candidate sequences, thereby significantly reducing computational complexity.
In another key step of beam search, generating $beam\_width \times top\_k$ candidate sequences from the existing $beam\_width$ sequences, there exists an important property: $log\_probs$ of each sequence are arranged in descending order. Leveraging this property, an optimized early termination mechanism can be adopted, where computations are halted prematurely if $log\_probs$ does not meet specific criteria, further enhancing computational efficiency. The specific operational procedure is as follows:

First, the existing $beam\_width$ sequences are accessed in order. For each accessed sequence, we create a small min-heap of size $beam\_width$, which is used to dynamically maintain the set of locally optimal elements currently selected.
Next, the subsequent sequences are traversed. During this process, if the subsequent $log\_probs$ of a sequence is smaller than the top element of the min-heap, it implies that, based on the current filtering criteria, the subsequent elements of this sequence cannot enter the current optimal set. At this point, the filtering operation for the current sequence can be terminated directly, and the next existing sequence processed. Conversely, if the subsequent $log\_probs$ is larger than the top element of the min-heap, the element is inserted into the min-heap, and the heap structure is adjusted according to the properties of the min-heap to maintain its order.
After the traversal of all relevant sequences is completed, the top elements are sequentially extracted from the min-heap. These elements, in the order of extraction, form the $beam\_width$ candidate sequences sorted in descending order. Through this procedure, the approach ensures the selection of high-quality candidate sequences while significantly reducing unnecessary computational overhead, thereby improving the operational efficiency of the beam search algorithm in practical applications.

\paragraph{Resource Reuse and Pre-allocation.}
During the beam search process, the $beam\_width$ is predetermined and fixed for each forward pass of the model, which implies that the computational resources required per forward propagation remains relatively constant. However, an excessively large $beam\_width$ can lead to significant wastage of both CPU computational and memory resources. To effectively mitigate this issue, xLLM incorporates a carefully designed resource reuse strategy. Specifically, during the generation of new candidate sequences, the system reuses resources previously occupied by older sequences without allocating new space for each candidate. This approach avoids the overhead associated with frequently creating new data structures. Based on the final search results, the system then updates the storage areas for the old sequences with the content of the newly generated ones upon completion of the beam search algorithm. By doing so, xLLM not only reduces memory usage but also minimizes the additional overhead on the CPU for resource management and allocation, significantly improving the efficiency of computational resource usage.

\subsubsection{Device-side Optimization}

\paragraph{Valid Item Filtering.}
In typical single-stage generative recommendation frameworks, the valid item filtering mechanism plays a crucial role. 
For example, OneRec \cite{onerec} framework uses an ordered combination of three token IDs to uniquely represent a valid product item ID. 
However, due to the vast number of possible token ID combinations, not all of them correspond to actual and valid items~\cite{tiger, sparse-meets-dense}. 
To ensure that the results generated by the model are all valid product items, the system generates a valid mask asynchronously during the forward pass of the model. This mask is based on a pre-constructed vocabulary of valid items, which is then element-wise added to the logits output by the model. Through this clever design, the logits corresponding to invalid token IDs are adjusted, ensuring that these invalid token IDs are almost never selected in subsequent calculations. This effectively filters out invalid token IDs, thereby guaranteeing the validity and accuracy of the final recommendation results.

\section{Evaluations}
\label{sec:evaluation}

In \textsection\ref{sec:main}, we evaluate the full implementation of our xLLM framework against several baseline inference systems, specifically vLLM-Ascend\footnote{\url{https://github.com/vllm-project/vllm-ascend}}(v0.10.rc1) and MindIE\footnote{\url{https://www.hiascend.com/en/software/mindie}}(v2.1.rc1), on Ascend 910B/910C~\cite{npu} instances across a range of scenarios. xLLM, MindIE, and vLLM-Ascend refer to the default deployments on Ascend 910B. Additionally, we denote the deployments of xLLM, MindIE and vLLM-Ascend on Ascend 910C as $\textrm{xLLM}^{\ddagger}$, $\textrm{MindIE}^{\ddagger}$ and $\textrm{vLLM-Ascend}^{\ddagger}$, respectively. The testing scenarios are organized into multiple categories reflecting diverse online serving applications, such as the JingYan AI chatbot, customer service assistants, merchant assistants, product understanding and marketing recommendation systems. Additionally, in \textsection\ref{sec:ablation}, we conduct a detailed ablation study to assess the individual contributions of each optimized module.

\begin{figure}[t]
  \centering \includegraphics[width=0.95\textwidth]{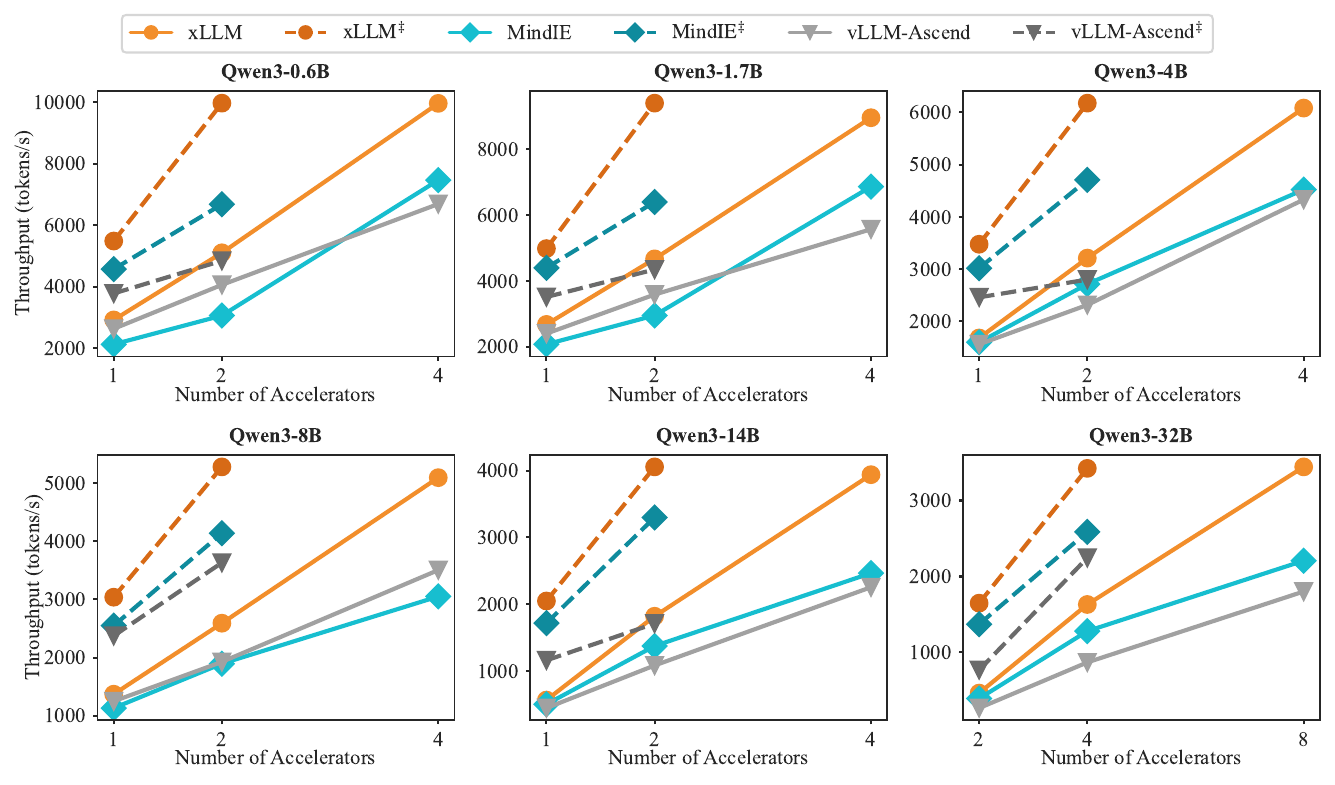} 
  \caption{Throughput comparison of LLM inference frameworks evaluated on Qwen3-series models using the ShareGPT dataset, under the constraint of \textit{TPOT=50ms} and \textit{input/output length=2048}.}
  \label{fig:benchmarkqw}
\end{figure}

\subsection{Main Results}
\label{sec:main}

This section benchmarks xLLM against mainstream inference frameworks to demonstrate its superior inference efficiency across diverse models, including Qwen2/3-series~\cite{qwen2, qwen3} and Deepseek~\cite{deepseek} models, and datasets. \textsection\ref{sec:benchmark} details a fair comparison using the ShareGPT\footnote{\url{https://huggingface.co/datasets/anon8231489123/ShareGPT_Vicuna_unfiltered/blob/main/ShareGPT_V3_unfiltered_cleaned_split.json}} dataset. To ensure an equitable comparison across all LLM inference frameworks, \textbf{the key feature of our experimental setup is that the input and output sequence lengths are fixed, while the request rate is dynamically adjusted to match the target SLO (e.g., TPOT) threshold for each framework.} We also evaluate this setup under different node configurations, including both single-node and multi-node setups with PD disaggregation. In \textsection\ref{sec:business_scenarios}, we evaluate xLLM on various real-world business scenarios from JD.com, where it is currently deployed, showcasing its performance under practical deployment conditions.

\subsubsection{Benchmarking Performance} \label{sec:benchmark}

\begin{figure}[t]
  \centering
  \includegraphics[width=1.0\textwidth]{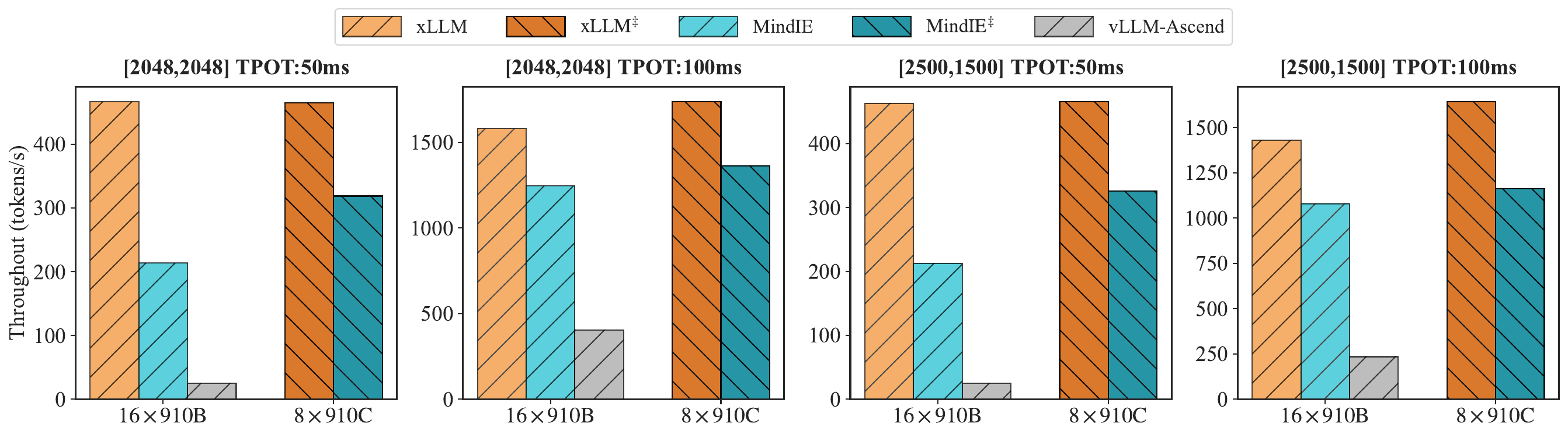} 
  \caption{Throughput comparison of LLM inference frameworks using DeepSeek-R1 on the ShareGPT dataset constrained by given TPOT and input/output length, where \textit{[2500, 1500]} means \textit{input length=2500} and \textit{output length=1500}. vLLM-Ascend on the 910C is excluded from the comparison, as its performance with Deepseek-R1 fails to meet the required TPOT threshold.}
  \label{fig:benchmarkds}
\end{figure}

\begin{table}[t]
\centering
\resizebox{\textwidth}{!}{
\begin{tabular}{c|c|c|c|c}
\toprule
\textbf{Method} & \textbf{Prompt Length} & \textbf{Output Length} & \textbf{Throughput (tokens/s)} & \textbf{Request Rate (req/s)}\\
\midrule
MindIE & 2048 & 2048 & 8476.44 & 4.14 \\
\midrule
xLLM & 2048 & 2048 & \textbf{11351.58} & \textbf{5.54} \\
\bottomrule
\end{tabular}
}
\vspace{2mm}
\caption{Comparison of DeepSeek-R1 with PD disaggregation on the ShareGPT dataset constrained by \textit{TPOT=100ms}.}
\label{table:benchmark_ds_pd}
\end{table}

\paragraph{Qwen3-series.} As shown in Figure~\ref{fig:benchmarkqw}, we comprehensively compares the throughput performance of four inference frameworks across the Qwen3-series (from 0.6B to 32B parameters) on the benchmark. 
All tests are conducted under uniform configuration with input/output lengths set to 2048 tokens and a TOPT constraint of 50 ms. 
The experimental results demonstrate that while all evaluated frameworks show improved throughput with additional accelerators, xLLM and its Ascend 910C implementation ($\textrm{xLLM}^{\ddagger}$) consistently deliver superior performance, confirming near-linear strong scalability across various model sizes. Specifically, xLLM achieves throughput improvements of up to 1.9$\times$ and 1.7$\times$ compared to vLLM-Ascend and MindIE, respectively. Similarly, $\textrm{xLLM}^{\ddagger}$ outperforms $\textrm{vLLM-Ascend}^{\ddagger}$ and $\textrm{MindIE}^{\ddagger}$ by up to 2.2$\times$ and 1.5$\times$, respectively.
Furthermore, $\textrm{xLLM}^{\ddagger}$ demonstrates steady performance improvements over the Ascend 910B-based xLLM in most scenarios, validating the software stack's effective utilization of the new hardware.

\paragraph{DeepSeek-R1.} Figure~\ref{fig:benchmarkds} presents the throughput performance of various inference frameworks for the DeepSeek-R1 model on the benchmark, using 16 accelerators for Ascend 910B and 8 accelerators for Ascend 910C, respectively. The results demonstrate that the proposed xLLM framework achieves exceptional throughput on the Ascend 910 series. Quantitatively, xLLM on Ascend 910B delivers an average throughput improvement of approximately 1.7$\times$ over MindIE and a significant 12$\times$ enhancement over vLLM-Ascend. Furthermore, $\textrm{xLLM}^{\ddagger}$ achieves an average throughput increase of approximately 1.4$\times$ compared to $\textrm{MindIE}^{\ddagger}$.

\paragraph{PD Disaggregation Settings.} Table~\ref{table:benchmark_ds_pd} benchmarks the inference performance of the MindIE and xLLM frameworks on the DeepSeek-R1 model using a PD disaggregation architecture. Under identical conditions with TPOT controlled at 100ms for 2048-length inputs/outputs, xLLM achieves approximately 34\% higher throughput (11,351.58 vs. 8,476.44 tokens/s) and request rate (5.54 vs. 4.14 req/s), marking a significant efficiency improvement.


\subsubsection{Business Serving Scenarios} \label{sec:business_scenarios}

\begin{figure}[t]
  \centering
  \includegraphics[width=1.0\textwidth]{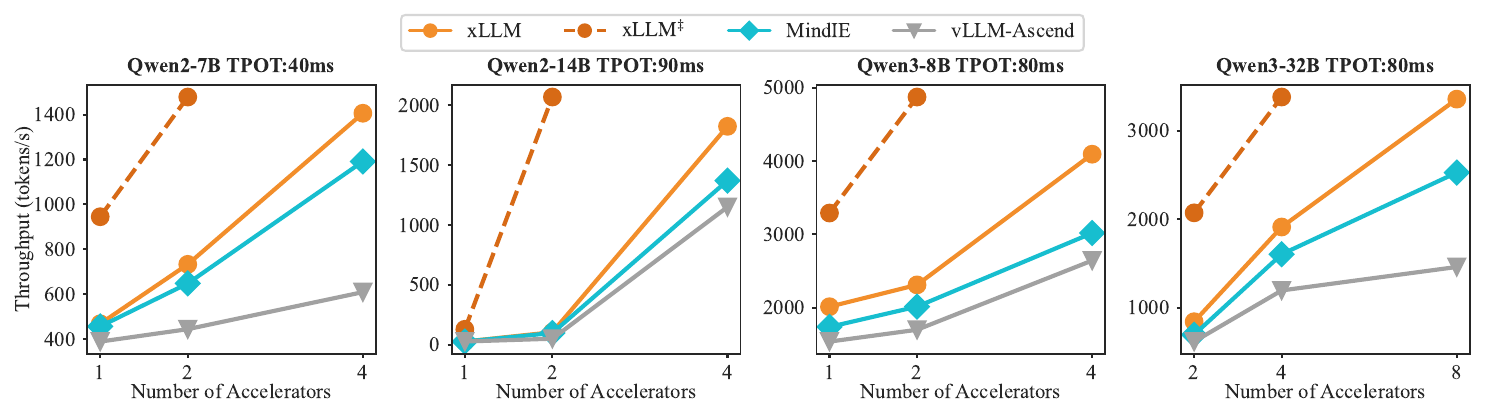} 
  \caption{Throughtput comparison of Qwen2-series and Qwen3-series models with various inference frameworks in the JingYan scenario.}
  \label{fig:jingyanqw}
\end{figure}

\paragraph{JingYan.} JingYan is an AI shopping assistant designed to help users discover new products, find inspiration, and get answers to their questions. The dataset for JingYan, consequently, consists of conversational logs between the model and its users, capturing these rich interactions. As illustrated in Figure~\ref{fig:jingyanqw}, we systematically evaluate the inference throughput of four frameworks for serving the Qwen2-series and Qwen3-series model in the JingYan scenario. The results demonstrate that both xLLM and $\textrm{xLLM}^{\ddagger}$ maintain superior throughput and exhibit better scaling efficiency across all model sizes compared to MindIE and vLLM-Ascend. For instance, when serving the Qwen3-8B model with 4 accelerators, xLLM delivers a throughput approximately 1.6 times that of vLLM-Ascend and significantly surpasses MindIE. The robust performance of xLLM on the 910B, and its enhanced results on the 910C hardware, highlight its effective adaptation to successive hardware generations. As shown in Table~\ref{table:jingyan_ds}, a similar trend is observed for the DeepSeek-V3 model, where xLLM achieves a throughput that is over 9 times greater than vLLM-Ascend and surpasses MindIE by 36\%.

\begin{table}[t]
\centering
\resizebox{\textwidth}{!}{
\begin{tabular}{c|c|c|c|c}
\toprule
\textbf{Method} & \textbf{Prompt Length} & \textbf{Output Length} & \textbf{Throughput (tokens/s)} & \textbf{Request Rate (req/s)}\\
\midrule
vLLM-Ascend & 6800 & 400 & 21.17 & 0.11 \\
\midrule
MindIE & 6800 & 400 & 144.40 & 0.67 \\
\midrule
xLLM & 6800 & 400 & \textbf{196.45} & \textbf{0.89} \\
\bottomrule
\end{tabular}
}
\vspace{2mm}
\caption{Comparison of Deepseek-V3 model with various frameworks in the JingYan scenario constrained by \textit{TPOT=80ms}.}
\label{table:jingyan_ds}
\end{table}

\begin{figure}[t]
  \centering
  \includegraphics[width=0.8\textwidth]{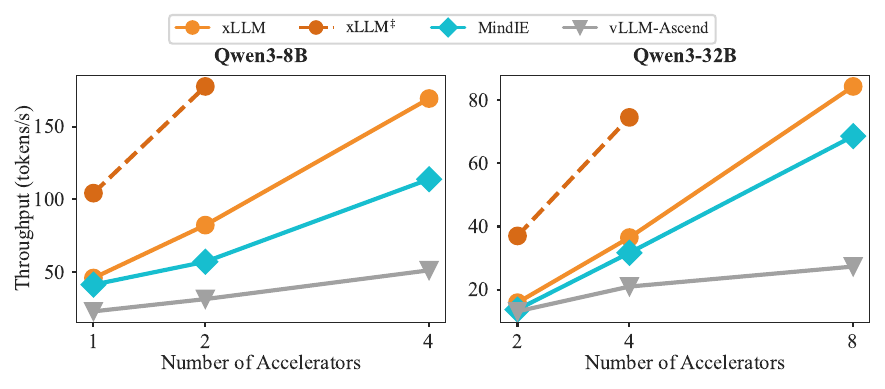} 
  \caption{Throughtput comparison of Qwen3-series models with various inference frameworks in the customer service scenario constrained by \textit{End-to-End latency(E2E)=10s}.}
  \label{fig:kefuqw}
\end{figure}

\paragraph{Customer Service.} The Customer Service dataset comprises the interactive dialogues between customers and support agents. Figure~\ref{fig:kefuqw} details the performance differences among inference frameworks for the Qwen-8B and Qwen-32B models in the customer service scenario. 
The results highlight that our xLLM, particularly the $\textrm{xLLM}^{\ddagger}$ running on Ascend 910C, provide significantly better throughput across all tested configurations. For instance, with Qwen3-32B on 8 accelerators, the throughput of xLLM is 3.1 and 1.2 times greater than that of vLLM-Ascend and MindIE, respectively.
It is important to note that the vLLM-Ascend framework shows a clear scaling bottleneck as the number of accelerators increases, whereas xLLM maintains near-linear efficiency scaling.
This validates the high efficiency and superiority of the xLLM framework in managing distributed inference for large-scale models.

\begin{figure}[t]
  \centering
  \includegraphics[width=1.0\textwidth]{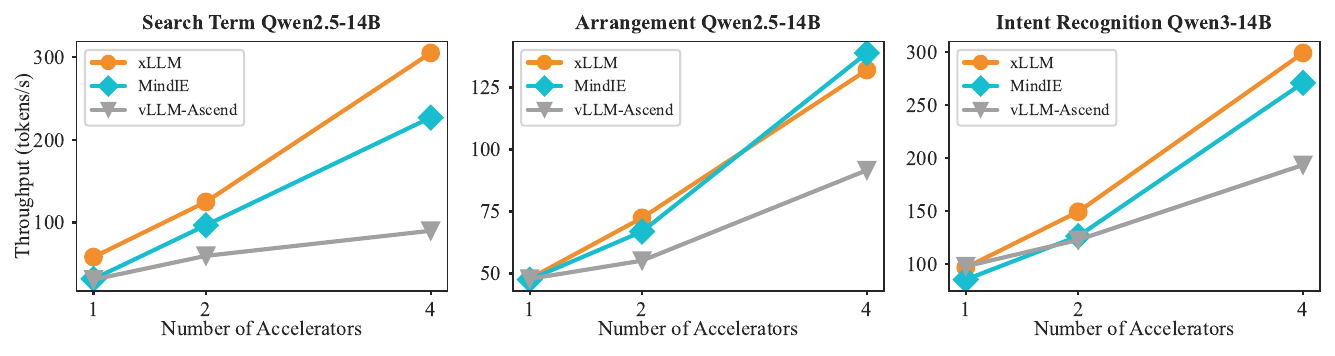} 
  \caption{Throughtput comparison of Qwen-series models with various inference frameworks in the merchant assistant scenario constrained by \textit{End-to-End latency(E2E)=1s}.}
  \label{fig:assistant}
\end{figure}

\paragraph{Merchant Assistant.}
Table~\ref{fig:assistant} benchmarks the throughput of various inference frameworks on three tasks (i.e., search terms, arrangement, intent recognition) within the merchant assistant scenario. The proposed xLLM framework achieves better or comparable performance to MindIE and demonstrates a significant lead over vLLM-Ascend. Specifically, for the search terms task with four accelerator cards, xLLM delivers 34\% higher throughput than MindIE and roughly 3.4$\times$ that of vLLM-Ascend.

\paragraph{Product Understanding.}
For the product understanding scenario, the throughput comparison of the Qwen2-7B model with several frameworks is shown in Table~\ref{table:understanding}. The experimental results indicate that xLLM outperforms MindIE and vLLM-Ascend by an average of 25\% and 56\%, respectively, across different accelerator card counts. Moreover, the superiority of xLLM scales with the number of cards, demonstrating its effective utilization of large-scale parallel computing resources and thereby offering a robust solution for high-performance LLM inference.

\begin{table}[t]
\centering
\resizebox{0.8\textwidth}{!}{
\begin{tabular}{c|c|c|c|c|c}
\toprule
\multirow{2}{*}{\textbf{Method}} & \multirow{2}{*}{\textbf{Prompt Length}} & \multirow{2}{*}{\textbf{Output Length}} & \multicolumn{3}{c}{\textbf{Throughput (tokens/s)}}
 \\
\cmidrule{4-6}
& & & \#accelerator=1 & \#accelerator=2 & \#accelerator=4 \\
\midrule
vLLM-Ascend & 1200 & 40 & 795.77 & 874.97 & 1272.52 \\
\midrule
MindIE & 1200 & 40 & 944.81 & 1051.44 & 1693.45 \\
\midrule
xLLM & 1200 & 40 & \textbf{1001.91} & \textbf{1323.90} & \textbf{2425.13} \\
\bottomrule
\end{tabular}
}
\vspace{2mm}
\caption{Throughtput comparison of Qwen2-7B model with various inference frameworks in the product understanding scenario.}
\label{table:understanding}
\end{table}

\begin{figure}[t]
  \centering
  \includegraphics[width=1.0\textwidth]{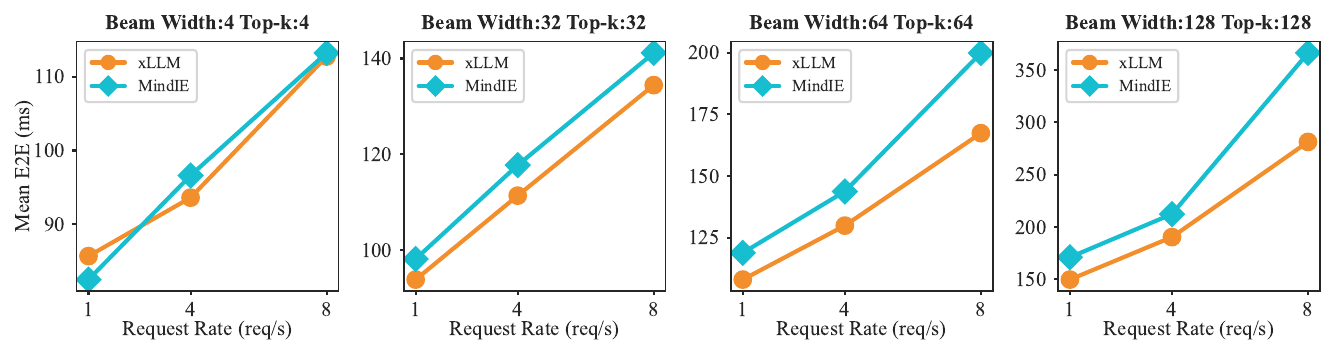} 
  \caption{Comparison of mean End-to-End latency(E2E) with various inference frameworks in the generative recommendation scenario. Since vLLM-Ascend does not support \textit{beam width>10}, the corresponding results are not plotted in the figure. Even with \textit{beam width=4} and \textit{request rate=1}, its mean E2E far exceeds that of our framework.}
  \label{fig:beam}
\end{figure}

\paragraph{Generative Recommendation.} As illustrated in Figure~\ref{fig:beam}, the evaluation results on the Qwen-8B model demonstrate that xLLM consistently achieves lower mean end-to-end latency than MindIE across various request rates and beam widths, except under very low load condition.
Notably, the performance advantage of xLLM becomes increasingly pronounced as the beam width (from 4 to 128) and the request rate escalate. For instance, under the most challenging scenario with a beam width of 128 and a request rate of 8, xLLM reduces latency by approximately 23\% relative to MindIE. 
This significant improvement validates that our xLLM's host-side and device-side optimizations effectively alleviate the computational bottlenecks in generative recommendation tasks, markedly enhancing inference efficiency and scalability under heavy loads.

\subsection{Ablation Study}
\label{sec:ablation}




\paragraph{Impact of MTP.}
As shown in Figure~\ref{fig:mtp}, under the configuration of 1500 input length and 2500 output length for the DeepSeek-R1 model, enabling Multi-Token Prediction (MTP) technology significantly optimizes inference performance. As max concurrency increases, the MTP-enabled version consistently exhibits lower TPOT compared to the baseline, indicating reduced generation latency. Meanwhile, its throughput is markedly higher than the non-MTP version, with particularly notable advantages beyond 32 concurrent requests. This demonstrates that MTP effectively enhances computational efficiency and system throughput under high concurrency conditions.

\begin{figure}[t]
  \centering
  \includegraphics[width=0.8\textwidth]{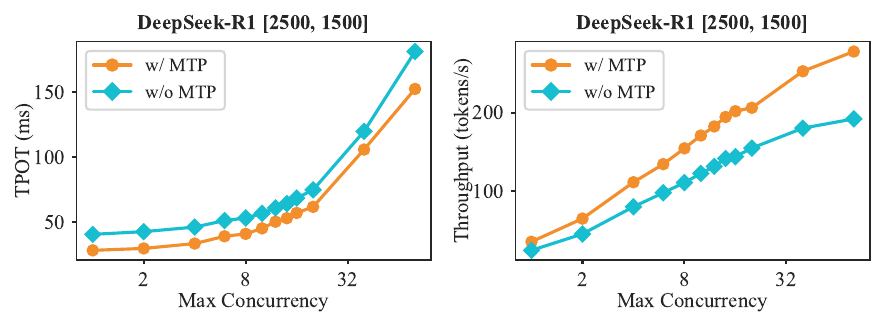} 
  \caption{Impact of MTP on the concurrent performance of DeepSeek-R1 model.}
  \label{fig:mtp}
\end{figure}

\begin{figure}[t]
  \centering
  \includegraphics[width=0.7\textwidth]{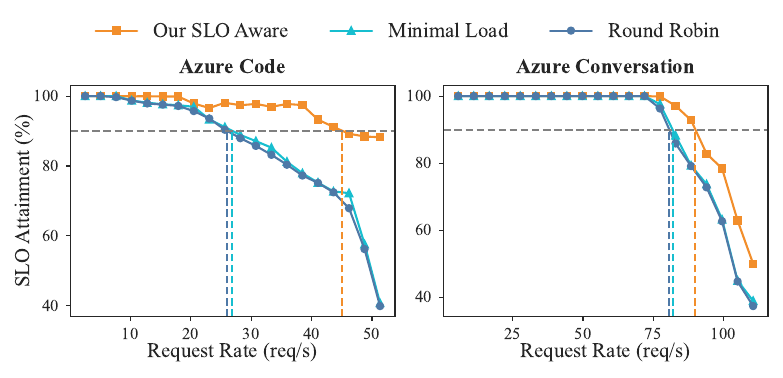} 
  \caption{Performance of Dynamic PD Disaggregation Policy with different scheduling strategies.}
  \label{fig:strategyArrow}
\end{figure}

\paragraph{Impact of Dynamic PD Disaggregation Scheduler Policy.}
We evaluate our proposed SLO-aware Dynamic PD Disaggregation Policy against the Minimal Load and Round Robin strategies, as illustrated in Figure~\ref{fig:strategyArrow}. On the Azure Code dataset characterized significant bursty traffic, our SLO-aware policy achieves a request serving rate 1.67 times that of the Minimal Load strategy. Meanwhile, relative to the Round Robin strategy, the Minimal Load strategy improves the SLO attainment by up to 4.3\%. For the Azure Conversation dataset with stable input/output length variations, our SLO-aware policy results in a 1.1 times higher request serving rate compared to the Minimal Load strategy. Although the Minimal Load strategy performs similarly to Round Robin, it still enhances the SLO attainment by up to 2.4\%. These results indicate that minimal-load scheduling is closer to the optimal scheduling strategy than round-robin approach,  while our adaptive instance scheduling delivers the best overall performance.

\begin{figure}[t]
  \centering
  \includegraphics[width=0.5\textwidth]{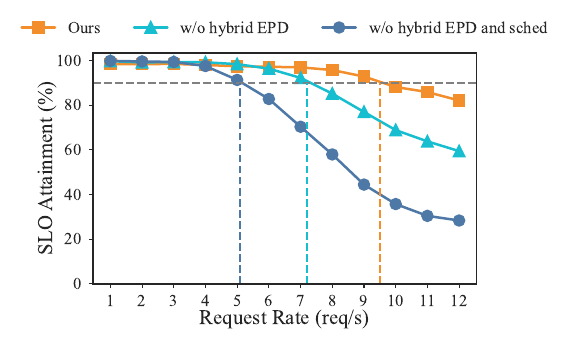} 
  \caption{Impact of Hybrid EPD Disaggregation Scheduler Policy.}
  \label{fig:epd}
\end{figure}

\begin{figure}[t]
  \centering
  \includegraphics[width=0.8\textwidth]{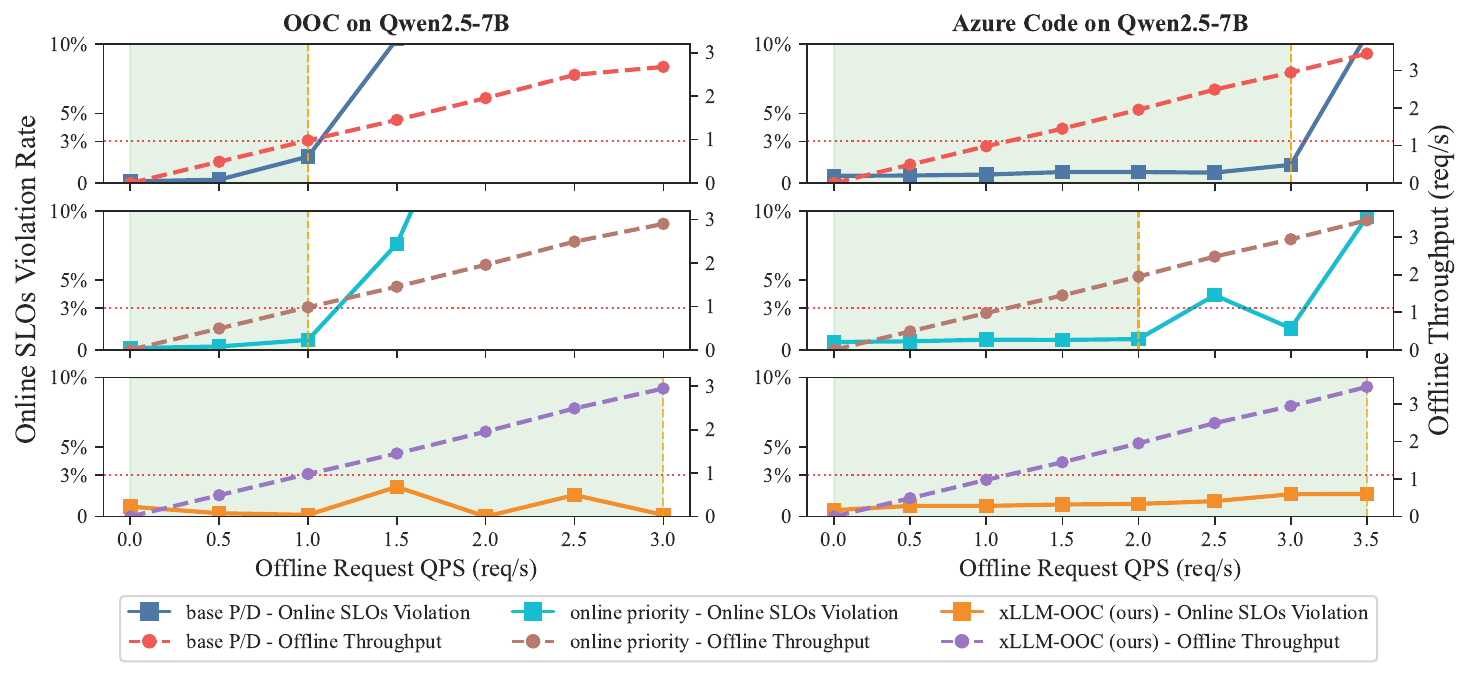} 
  \caption{Impact of our Online-Offline Co-location Scheduler Policy.}
  \label{fig:ooc_exp}
\end{figure}

\paragraph{Impact of Hybrid EPD Disaggregation Scheduler Policy.}
Figure~\ref{fig:epd} demonstrates the effectiveness of the proposed Hybrid EPD Disaggregation Policy on the TextCaps dataset, specifically in controlling TPOT and improving the SLO attainment rate. Under the configuration of 8 general-purpose inference instances, the removal of the hybrid EPD disaggregation method results in a drop in goodput from 9.5 req/s to 7.2 req/s, and a subsequent additional removal of stage-level scheduling policy further decreases the goodput to 5.1 req/s. This indicates that a well-designed disaggregation strategy can effectively mitigate inter-stage interference, while our stage-level scheduling policy contributes to finer control over the execution time of each batch. The integration of both components significantly enhances the overall performance and stability of the system, particularly under high-concurrency scenarios.

\paragraph{Impact of Online-Offline Co-location Scheduler Policy.}
As shown in Figure~\ref{fig:ooc_exp}, we evaluated three scheduling strategies, namely the baseline P/D, online priority and our proposed Online-Offline Co-location Scheduler Policy (denoted as xLLM-OOC), to assess the maximum achievable throughput for offline requests without violating the SLO of online requests. The green shaded region in the figure indicates the sustainable range of offline request throughput that remains within the acceptable online SLO violation threshold. When the offline query-per-second (QPS) exceeds a certain level, both the baseline P/D and online priority strategies lead to a sharp increase in the online SLO violation rate, indicating interference caused by the offline workload. In contrast, xLLM-OOC maintains stable SLO compliance even as offline QPS continues to rise. On our proprietary dataset, xLLM-OOC achieves a throughput that is three times higher than the other two methods. Furthermore, it demonstrates improvements of 75\% and 17\% over online priority and baseline P/D, respectively, on the Azure Code dataset.

\begin{table}[t]
\centering
\resizebox{\textwidth}{!}{
\begin{tabular}{c|c|c|c|c}
\toprule
\textbf{Model} & \textbf{Prompt Length} & \textbf{Output Length} & \textbf{Async Scheduling} & \textbf{Throughput (tokens/s)} \\
\midrule
\multirow{2}{*}{DS-Distill-Qwen-1.5B} & 1000 & 1000 & \off & 8709.90 \\

 & 1000 & 1000 & \on & \textbf{10223.15} \\
\midrule
\multirow{2}{*}{DS-Distill-Qwen-7B} & 1000 & 1000 & \off & 3183.68 \\

 & 1000 & 1000 & \on & \textbf{3201.83} \\
\midrule
\multirow{2}{*}{DS-Distill-Qwen-14B} & 1000 & 1000 & \off & 1472.22 \\

 & 1000 & 1000 & \on & \textbf{1527.23} \\
\midrule
\multirow{2}{*}{DS-Distill-Qwen-32B} & 1000 & 1000 & \off & 1415.20 \\

 & 1000 & 1000 & \on & \textbf{1508.76} \\
\bottomrule
\end{tabular}
}
\vspace{2mm}
\caption{Impact of asynchronous scheduling mechanism in Multi-layer Pipeline Execution.}
\label{table:async}
\end{table}

\begin{table}[t]
\centering
\begin{tabular}{c|c|c}
\toprule
\textbf{Metric} & \textbf{Single-Stream (ms)} & \textbf{Dual-Stream (ms)} \\
\midrule
Total Communication Time & 9.3 & 12.4 \\
Overlapped Communication Ratio & - & 80\% \\
Exposed Communication Time & 9.3 & 2.5 \\
Total Computation Time & 13.0 & 17.0 \\
Reduced Time per Layer & - & 2.8 \\
Total Reduced Time (61 layers) & - & 172.0 \\
\bottomrule
\end{tabular}
\vspace{2mm}
\caption{Communication and computation overheads in a single decoder layer of DeepSeek-R1 using Multi-layer Pipeline Execution.}
\label{tab:overlap_results}
\end{table}

\begin{table}[t]
\centering
\resizebox{\textwidth}{!}{
\begin{tabular}{c|c|c|c|c|c}
\toprule
\textbf{Model} & \textbf{Prompt Length} & \textbf{Output Length} & \textbf{Adaptive Graph Mode} & \textbf{Throughput (tokens/s)} & \textbf{Mean TPOT (ms)}\\
\midrule
\multirow{2}{*}{Qwen3-1.7B} & 2048 & 2048 & \off & 2385.58 & 39.27 \\

 & 2048 & 2048 & \on & \textbf{3038.88} & \textbf{30.63} \\
\midrule
\multirow{2}{*}{Qwen3-4B} & 2048 & 2048 & \off & 1539.93 & 55.44 \\

 & 2048 & 2048 & \on & \textbf{1671.39} & \textbf{50.58} \\
\bottomrule
\end{tabular}
}
\vspace{2mm}
\caption{Impact of Adaptive Graph Mode.}
\label{table:graphexp}
\end{table}

\paragraph{Impact of Multi-layer Pipeline Execution.}
As evidenced by the results in Table~\ref{table:async}, our proposed asynchronous scheduling mechanism delivers consistent throughput improvements across all evaluated model scales. The most substantial relative gain of 17.4\% is observed for the 1.5B parameter model, highlighting the method's particular efficacy for smaller architectures where scheduling overhead constitutes a larger portion of total computation time. While relative improvements moderate for larger models (reaching 0.6\% for 7B, 3.7\% for 14B, and 6.6\% for 32B), all configurations exhibit statistically significant absolute gains. These results robustly validate that our method successfully masks scheduling latency and eliminates computational bubbles, through leveraging placeholder tokens to decouple and overlap CPU scheduling with NPU execution.

We further assess the effectiveness of the proposed dual-stream architecture for the DeepSeek-R1 model in Table~\ref{tab:overlap_results}. Experimental results from a single decoder layer show that the total communication time increases to 12.4ms in dual-stream mode, up from 9.3ms in single-stream mode. However, the computation-communication overlap mechanism successfully hides 80\% of the communication time, which brings the exposed communication time down to just 2.5ms, saving 6.8ms per layer. Despite introducing a computational overhead of 4ms for each layer, the dual-stream strategy yields a net performance gain of 172.0ms across the entire 61-layer model, clearly illustrating the capability of our scheduling strategy in real-world workloads.

\paragraph{Impact of Adaptive Graph Mode.}
To validate the effectiveness of the Adaptive Graph Mode optimization technique, we conduct controlled experiments on the Qwen3-1.7B and Qwen3-4B models. As shown in Table~\ref{table:graphexp}, with both the prompt and output lengths set to 2048 tokens, enabling Adaptive Graph Mode result in significant performance gains for both models. The throughput of Qwen3-1.7B increases from 2385.58 to 3038.88 tokens/s (a 27.4\% improvement), while its mean TPOT decreases by 22.0\%. Correspondingly, Qwen3-4B achieved an 8.5\% increase in throughput and an 8.8\% reduction in latency. These results robustly demonstrate the effectiveness of Adaptive Graph Mode as a general-purpose inference acceleration technique, with more pronounced optimization effects observed on the model with a smaller parameter size.

\paragraph{Impact of Hierarchical DP Load Balancing.}
Our proposed hierarchical DP load balancing scheme is projected to increase total throughput by 5\%. The kernel-level optimization yields the most significant latency savings; for example, for an ultra-long 32k-token request, reordering and splitting can reduce a single core's computational load from 32k to 1.3k tokens, saving approximately 800 microseconds. In contrast, the latency savings from inter-DP group migration are more modest. Even after balancing a 20k-token difference, the total time saved over 61 Transformer layers is only about 600 microseconds. This indicates that the greatest performance bottleneck and optimization potential lie within the computation units themselves.

\section{Future Work}

Although proposed xLLM framework has demonstrated potential in enhancing LLM inference efficiency and reducing operational cost, achieving truly efficient and inclusive intelligent computing infrastructure requires deeper collaborative innovation across \textbf{Hardware}, \textbf{Model}, and \textbf{Application} Layers. 
Our future research will advance xLLM from a high-performance \textit{Inference Engine} toward a comprehensive \textit{Operating System for AI} that supports next-generation intelligent applications. We outline our planned efforts in three key directions:

\subsection{Fostering an Open and Diverse Hardware Ecosystem}
Current AI infrastructure often suffers from tight coupling with specific hardware architectures, introducing computational stability risks and limiting efficient deployment in heterogeneous environments such as edge computing. To address this, the first pillar aims to cultivate an open and diverse hardware ecosystem through the following initiatives:

\paragraph{Unified Hardware Abstraction Layer.}
We will develop a high-performance, hardware-agnostic runtime abstraction layer that encapsulates instruction sets and memory architectures across computational units—from domestic cloud chips to edge-side accelerators. This layer will provide unified operator interfaces and memory management APIs, enabling seamless migration and efficient execution across heterogeneous hardware without code modifications, thereby significantly reducing integration barriers for emerging hardware.

\paragraph{Software-Hardware Co-Design and Ecosystem Synergy.}
We will collaborate with hardware partners to define more efficient and open interface standards. This co-design approach will not only drive hardware innovation from the software perspective but also help establish a self-sustaining and competitive computational supply ecosystem, ultimately providing users with optimized choices across cost, performance, and security considerations.

\subsection{Cultivating a Vibrant and Responsive Model Ecosystem}
The system's core value derives from its supported model diversity and integration efficiency. Our future efforts will focus on building an inclusive and agile model ecosystem through:

\paragraph{Multi-Scenario Support.}
We will extend the platform's optimization and deployment capabilities beyond large language models to systematically support various generative AI scenarios, including generative recommendation, text-to-image, and text-to-video generation. We will optimize execution engines, request scheduling, and memory management specifically for different workloads to ensure optimal performance across diverse applications.

\paragraph{``Zero-Day'' Model Integration.}
To accommodate rapidly evolving model architectures, we will implement a unified framework combining model graph representation, an extensible operator library, and automated compilation optimization. This will enable rapid ``zero-day'' integration of newly released models, reducing deployment cycles from weeks to hours.

\subsection{Evolving into an AI-Native Application Framework}

To democratize AI adoption and accelerate value delivery, we will evolve the system from an inference engine into a full-stack AI-native application framework by:

\paragraph{Framework-Native AI Middleware.}
We will design high-level, framework-native APIs and abstractions that package complex distributed inference, multi-model orchestration, and stateful session management capabilities into out-of-the-box middleware services. This will enable application developers to build sophisticated AI applications (e.g., multimodal AI agents) without managing underlying infrastructure complexities.

\paragraph{Rapid Application Integration and Deployment.}
Building upon this AI-native framework, we will deliver a comprehensive toolchain --- including encapsulated SDKs, application templates, and integrated CI/CD pipelines --- to streamline development and deployment processes. Our goal is to enable application teams to integrate and deploy AI services within hours rather than weeks, significantly enhancing business innovation agility and fully bridging the ``last mile'' from model development to value creation.

\section{Conclusion}


We introduce \textbf{xLLM}, an intelligent and efficient LLM inference framework, featuring a service-engine decoupled architecture. The framework consists of two main components: (1) \textbf{xLLM-Service}, a versatile service layer designed for efficient instance and cluster management as well as request scheduling. It incorporates unified elastic scheduling for co-located online and offline requests to maximize cluster utilization, a workload-adaptive PD disaggregation architecture for SLO-aware dynamic instance scheduling, a novel Encode-Prefill-Decode (EPD) disaggregation mechanism for multimodal requests, 
a global KV Cache manager for efficient memory management including KV Cache upload and offload and a distributed fault-tolerance design to ensure high availability. (2) \textbf{xLLM-Engine}, a high-performance inference engine optimized for accelerating LLM inference across various AI accelerators. xLLM-Engine employs full-stack multi-layer execution pipeline optimizations through asynchronizing CPU-side scheduling and AI accelerator-side model forwarding to minimize computational bubbles, utilizing dual-stream parallelism of micro-batches to overlap computation with all-to-all communication, and further overlapping various AI computation units at the operator level. xLLM-Engine also implements an adaptive graph mode which pre-compiles kernel sequences, memory operations, and synchronization into a single computation graph to drastically reduce launch overhead and accelerator idle time. Additionally, it adopts a "logically contiguous, physically discrete" KV Cache storage strategy via proposed xTensor Memory Management, which resolves the tension between memory contiguity and dynamic allocation. To further boost hardware utilization, xLLM-Engine integrates algorithmic enhancements such as asynchronous pipelined adaptive speculative decoding, KV cache-aware scheduling, reactive inter-DP group workload migration for dynamic load balancing, and dynamic MoE load balancing based on real-time expert workload statistics and transparent weight updates. We have further extended xLLM to emerging generative recommendation scenarios, improving both recommendation accuracy and efficiency.

\quad Extensive experiments demonstrate that xLLM achieves a consistent improvement compared to leading inference systems such as MindIE and vLLM-Ascend, evaluated across mainstream Qwen-series and Deepseek-series models as well as public and real-world industrial datasets. In particular, xLLM outperforms MindIE by up to 1.7$\times$ and vLLM-Ascend by up to 2.2$\times$ in throughput. Comprehensive ablation studies further validate the effectiveness of key components, including the proposed scheduling modules, multi-layer execution pipeline, optimized model tensor parallelism, adaptive graph mode, and others.

\quad By releasing the xLLM framework as an open-source project, we intend to stimulate further innovation in developing robust, enterprise-scale inference solutions, optimizing performance on a diverse range of AI accelerators, and creating tightly integrated service and engine architectures for next-generation AI applications.
\newpage
{
    \small
    \bibliographystyle{unsrt}
    \bibliography{neurips_2023}

@String(IJCAI = {IJCAI})

@inproceedings{kwon2023efficient,
  title={Efficient Memory Management for Large Language Model Serving with PagedAttention},
  author={Woosuk Kwon and Zhuohan Li and Siyuan Zhuang and Ying Sheng and Lianmin Zheng and Cody Hao Yu and Joseph E. Gonzalez and Hao Zhang and Ion Stoica},
  booktitle={Proceedings of the ACM SIGOPS 29th Symposium on Operating Systems Principles},
  year={2023}
}

@misc{fastertransformer2023,
  title={FasterTransformer},
  author={NVIDIA},
  year={2023},
  howpublished={\url{https://github.com/NVIDIA/FasterTransformer}},
  note={Transformer related optimization, including BERT, GPT}
}

@inproceedings{qin2025mooncake,
  title={Mooncake: Trading more storage for less computation—a $\{$KVCache-centric$\}$ architecture for serving $\{$LLM$\}$ chatbot},
  author={Qin, Ruoyu and Li, Zheming and He, Weiran and Cui, Jialei and Ren, Feng and Zhang, Mingxing and Wu, Yongwei and Zheng, Weimin and Xu, Xinran},
  booktitle={23rd USENIX Conference on File and Storage Technologies (FAST 25)},
  pages={155--170},
  year={2025}
}

@misc{etcd,
  title={etcd: A distributed, reliable key-value store for the most critical data of a distributed system},
  author={{etcd Authors}},
  howpublished={\url{https://etcd.io}}
}

@article{dao2022flashattention,
  title={Flashattention: Fast and memory-efficient exact attention with io-awareness},
  author={Dao, Tri and Fu, Dan and Ermon, Stefano and Rudra, Atri and R{\'e}, Christopher},
  journal={Advances in neural information processing systems},
  volume={35},
  pages={16344--16359},
  year={2022}
}

@article{zhai2024actions,
  title={Actions speak louder than words: Trillion-parameter sequential transducers for generative recommendations},
  author={Zhai, Jiaqi and Liao, Lucy and Liu, Xing and Wang, Yueming and Li, Rui and Cao, Xuan and Gao, Leon and Gong, Zhaojie and Gu, Fangda and He, Michael and others},
  journal={arXiv preprint arXiv:2402.17152},
  year={2024}
}

@inproceedings{wang2024eager,
  title={Eager: Two-stream generative recommender with behavior-semantic collaboration},
  author={Wang, Ye and Xun, Jiahao and Hong, Minjie and Zhu, Jieming and Jin, Tao and Lin, Wang and Li, Haoyuan and Li, Linjun and Xia, Yan and Zhao, Zhou and others},
  booktitle={Proceedings of the 30th ACM SIGKDD Conference on Knowledge Discovery and Data Mining},
  pages={3245--3254},
  year={2024}
}

@article{gpt,
  title={Gpt-4 technical report},
  author={Achiam, Josh and Adler, Steven and Agarwal, Sandhini and Ahmad, Lama and Akkaya, Ilge and Aleman, Florencia Leoni and Almeida, Diogo and Altenschmidt, Janko and Altman, Sam and Anadkat, Shyamal and others},
  journal={arXiv preprint arXiv:2303.08774},
  year={2023}
}

@article{deepseek,
  title={Deepseek-v3 technical report},
  author={Liu, Aixin and Feng, Bei and Xue, Bing and Wang, Bingxuan and Wu, Bochao and Lu, Chengda and Zhao, Chenggang and Deng, Chengqi and Zhang, Chenyu and Ruan, Chong and others},
  journal={arXiv preprint arXiv:2412.19437},
  year={2024}
}

@article{llama,
  title={The llama 3 herd of models},
  author={Dubey, Abhimanyu and Jauhri, Abhinav and Pandey, Abhinav and Kadian, Abhishek and Al-Dahle, Ahmad and Letman, Aiesha and Mathur, Akhil and Schelten, Alan and Yang, Amy and Fan, Angela and others},
  journal={arXiv e-prints},
  pages={arXiv--2407},
  year={2024}
}

@inproceedings{customer-service,
  title={AntProphet: an Intention Mining System behind Alipay's Intelligent Customer Service Bot.},
  author={Chen, Cen and Zhang, Xiaolu and Ju, Sheng and Fu, Chilin and Tang, Caizhi and Zhou, Jun and Li, Xiaolong},
  booktitle={IJCAI},
  volume={8},
  pages={6497--6499},
  year={2019}
}

@article{recommendation,
  title={Generative recommendation: Towards next-generation recommender paradigm},
  author={Wang, Wenjie and Lin, Xinyu and Feng, Fuli and He, Xiangnan and Chua, Tat-Seng},
  journal={arXiv preprint arXiv:2304.03516},
  year={2023}
}

@inproceedings{content-generation,
  title={Llm-based interaction for content generation: A case study on the perception of employees in an it department},
  author={Agossah, Alexandre and Krupa, Fr{\'e}d{\'e}rique and Perreira Da Silva, Matthieu and Le Callet, Patrick},
  booktitle={Proceedings of the 2023 ACM International Conference on Interactive Media Experiences},
  pages={237--241},
  year={2023}
}

@misc{vllm,
  title = {vLLM},
  year = {2025},
  howpublished = {\url{https://github.com/vllm-project/vllm}},
}

@misc{sglang,
  title = {SGLang},
  year = {2025},
  howpublished = {\url{https://github.com/sgl-project/sglang}},
}

@misc{tensorrt,
  title = {TensorRT},
  year = {2025},
  howpublished = {\url{https://github.com/NVIDIA/TensorRT}},
}

@inproceedings{tidal-character,
  title={Elasticflow: An elastic serverless training platform for distributed deep learning},
  author={Gu, Diandian and Zhao, Yihao and Zhong, Yinmin and Xiong, Yifan and Han, Zhenhua and Cheng, Peng and Yang, Fan and Huang, Gang and Jin, Xin and Liu, Xuanzhe},
  booktitle={Proceedings of the 28th ACM International Conference on Architectural Support for Programming Languages and Operating Systems, Volume 2},
  pages={266--280},
  year={2023}
}

@inproceedings{distserve,
  title={$\{$DistServe$\}$: Disaggregating prefill and decoding for goodput-optimized large language model serving},
  author={Zhong, Yinmin and Liu, Shengyu and Chen, Junda and Hu, Jianbo and Zhu, Yibo and Liu, Xuanzhe and Jin, Xin and Zhang, Hao},
  booktitle={18th USENIX Symposium on Operating Systems Design and Implementation (OSDI 24)},
  pages={193--210},
  year={2024}
}

@misc{dynamo,
  title = {Dynamo},
  year = {2025},
  howpublished = {\url{https://github.com/ai-dynamo/dynamo}},
}

@article{multimodal1,
  title={Aim: Adaptive inference of multi-modal llms via token merging and pruning},
  author={Zhong, Yiwu and Liu, Zhuoming and Li, Yin and Wang, Liwei},
  journal={arXiv preprint arXiv:2412.03248},
  year={2024}
}

@article{multimodal2,
  title={Cosyvoice: A scalable multilingual zero-shot text-to-speech synthesizer based on supervised semantic tokens},
  author={Du, Zhihao and Chen, Qian and Zhang, Shiliang and Hu, Kai and Lu, Heng and Yang, Yexin and Hu, Hangrui and Zheng, Siqi and Gu, Yue and Ma, Ziyang and others},
  journal={arXiv preprint arXiv:2407.05407},
  year={2024}
}

@article{npu,
  title={Serving Large Language Models on Huawei CloudMatrix384},
  author={Zuo, Pengfei and Lin, Huimin and Deng, Junbo and Zou, Nan and Yang, Xingkun and Diao, Yingyu and Gao, Weifeng and Xu, Ke and Chen, Zhangyu and Lu, Shirui and others},
  journal={arXiv preprint arXiv:2506.12708},
  year={2025}
}

@inproceedings{tpu,
  title={Tpu v4: An optically reconfigurable supercomputer for machine learning with hardware support for embeddings},
  author={Jouppi, Norm and Kurian, George and Li, Sheng and Ma, Peter and Nagarajan, Rahul and Nai, Lifeng and Patil, Nishant and Subramanian, Suvinay and Swing, Andy and Towles, Brian and others},
  booktitle={Proceedings of the 50th annual international symposium on computer architecture},
  pages={1--14},
  year={2023}
}

@article{gshard,
  title={Gshard: Scaling giant models with conditional computation and automatic sharding},
  author={Lepikhin, Dmitry and Lee, HyoukJoong and Xu, Yuanzhong and Chen, Dehao and Firat, Orhan and Huang, Yanping and Krikun, Maxim and Shazeer, Noam and Chen, Zhifeng},
  journal={arXiv preprint arXiv:2006.16668},
  year={2020}
}

@inproceedings{infer-metric,
  title={Splitwise: Efficient generative llm inference using phase splitting},
  author={Patel, Pratyush and Choukse, Esha and Zhang, Chaojie and Shah, Aashaka and Goiri, {\'I}{\~n}igo and Maleki, Saeed and Bianchini, Ricardo},
  booktitle={2024 ACM/IEEE 51st Annual International Symposium on Computer Architecture (ISCA)},
  pages={118--132},
  year={2024},
  organization={IEEE}
}

@misc{node-fault,
      title={Robust LLM Training Infrastructure at ByteDance}, 
      author={Borui Wan and Gaohong Liu and Zuquan Song and Jun Wang and Yun Zhang and Guangming Sheng and Shuguang Wang and Houmin Wei and Chenyuan Wang and Weiqiang Lou and Xi Yang and Mofan Zhang and Kaihua Jiang and Cheng Ren and Xiaoyun Zhi and Menghan Yu and Zhe Nan and Zhuolin Zheng and Baoquan Zhong and Qinlong Wang and Huan Yu and Jinxin Chi and Wang Zhang and Yuhan Li and Zixian Du and Sida Zhao and Yongqiang Zhang and Jingzhe Tang and Zherui Liu and Chuan Wu and Yanghua Peng and Haibin Lin and Wencong Xiao and Xin Liu and Liang Xiang},
      year={2025},
      eprint={2509.16293},
      archivePrefix={arXiv},
      primaryClass={cs.LG},
      url={https://arxiv.org/abs/2509.16293}, 
}

@article{chunk-prefill,
  title={Efficient LLM Inference via Chunked Prefills},
  author={Agrawal, Arney and Kedia, Nitin and Panwar, Ashish and Mohan, Jayashree and Kwatra, Nipun and Gulavani, Bhargav S and Tumanov, Alexey and Ramjee, Ramachandran},
  journal={ACM SIGOPS Operating Systems Review},
  volume={59},
  number={1},
  pages={9--16},
  year={2025},
  publisher={ACM New York, NY, USA}
}

@misc{vllm-v1-schedule,
  title = {vLLM-V1-Scheduler},
  year = {2025},
  howpublished = {\url{https://github.com/vllm-project/vllm/issues/8779}},
}

@misc{tgi,
  title = {TGI},
  year = {2025},
  howpublished = {\url{https://github.com/huggingface/text-generation-inference}},
}

@misc{lina,
      title={Accelerating Distributed MoE Training and Inference with Lina}, 
      author={Jiamin Li and Yimin Jiang and Yibo Zhu and Cong Wang and Hong Xu},
      year={2024},
      eprint={2210.17223},
      archivePrefix={arXiv},
      primaryClass={cs.DC},
      url={https://arxiv.org/abs/2210.17223}, 
}

@article{tetriinfer,
  title={Inference without interference: Disaggregate llm inference for mixed downstream workloads},
  author={Hu, Cunchen and Huang, Heyang and Xu, Liangliang and Chen, Xusheng and Xu, Jiang and Chen, Shuang and Feng, Hao and Wang, Chenxi and Wang, Sa and Bao, Yungang and others},
  journal={arXiv preprint arXiv:2401.11181},
  year={2024}
}

@inproceedings{loongserve,
  title={Loongserve: Efficiently serving long-context large language models with elastic sequence parallelism},
  author={Wu, Bingyang and Liu, Shengyu and Zhong, Yinmin and Sun, Peng and Liu, Xuanzhe and Jin, Xin},
  booktitle={Proceedings of the ACM SIGOPS 30th Symposium on Operating Systems Principles},
  pages={640--654},
  year={2024}
}

@inproceedings{fapes,
  title={FaPES: Enabling Efficient Elastic Scaling for Serverless Machine Learning Platforms},
  author={Zhao, Xiaoyang and Yang, Siran and Wang, Jiamang and Diao, Lansong and Qu, Lin and Wu, Chuan},
  booktitle={Proceedings of the 2024 ACM Symposium on Cloud Computing},
  pages={443--459},
  year={2024}
}

@article{expertflow,
  title={Expertflow: Optimized expert activation and token allocation for efficient mixture-of-experts inference},
  author={He, Xin and Zhang, Shunkang and Wang, Yuxin and Yin, Haiyan and Zeng, Zihao and Shi, Shaohuai and Tang, Zhenheng and Chu, Xiaowen and Tsang, Ivor and Soon, Ong Yew},
  journal={arXiv preprint arXiv:2410.17954},
  year={2024}
}

@inproceedings{fastermoe,
  title={Fastermoe: modeling and optimizing training of large-scale dynamic pre-trained models},
  author={He, Jiaao and Zhai, Jidong and Antunes, Tiago and Wang, Haojie and Luo, Fuwen and Shi, Shangfeng and Li, Qin},
  booktitle={Proceedings of the 27th ACM SIGPLAN Symposium on Principles and Practice of Parallel Programming},
  pages={120--134},
  year={2022}
}

@article{mtp,
  title={Eagle-2: Faster inference of language models with dynamic draft trees},
  author={Li, Yuhui and Wei, Fangyun and Zhang, Chao and Zhang, Hongyang},
  journal={arXiv preprint arXiv:2406.16858},
  year={2024}
}

@article{deepboot,
  title={DeepBoot: Dynamic scheduling system for training and inference deep learning tasks in GPU cluster},
  author={Chen, Zhenqian and Zhao, Xinkui and Zhi, Chen and Yin, Jianwei},
  journal={IEEE transactions on parallel and distributed systems},
  volume={34},
  number={9},
  pages={2553--2567},
  year={2023},
  publisher={IEEE}
}

@article{roofline,
  title={Roofline: an insightful visual performance model for multicore architectures},
  author={Williams, Samuel and Waterman, Andrew and Patterson, David},
  journal={Communications of the ACM},
  volume={52},
  number={4},
  pages={65--76},
  year={2009},
  publisher={ACM New York, NY, USA}
}

@article{pd-serve,
  title={P/d-serve: Serving disaggregated large language model at scale},
  author={Jin, Yibo and Wang, Tao and Lin, Huimin and Song, Mingyang and Li, Peiyang and Ma, Yipeng and Shan, Yicheng and Yuan, Zhengfan and Li, Cailong and Sun, Yajing and others},
  journal={arXiv preprint arXiv:2408.08147},
  year={2024}
}

@inproceedings{p-compute,
  title={Orca: A distributed serving system for $\{$Transformer-Based$\}$ generative models},
  author={Yu, Gyeong-In and Jeong, Joo Seong and Kim, Geon-Woo and Kim, Soojeong and Chun, Byung-Gon},
  booktitle={16th USENIX Symposium on Operating Systems Design and Implementation (OSDI 22)},
  pages={521--538},
  year={2022}
}

@inproceedings{d-memory,
  title={Efficient memory management for large language model serving with pagedattention},
  author={Kwon, Woosuk and Li, Zhuohan and Zhuang, Siyuan and Sheng, Ying and Zheng, Lianmin and Yu, Cody Hao and Gonzalez, Joseph and Zhang, Hao and Stoica, Ion},
  booktitle={Proceedings of the 29th symposium on operating systems principles},
  pages={611--626},
  year={2023}
}

@inproceedings{bytecheckpoint,
  title={$\{$ByteCheckpoint$\}$: A Unified Checkpointing System for Large Foundation Model Development},
  author={Wan, Borui and Han, Mingji and Sheng, Yiyao and Peng, Yanghua and Lin, Haibin and Zhang, Mofan and Lai, Zhichao and Yu, Menghan and Zhang, Junda and Song, Zuquan and others},
  booktitle={22nd USENIX Symposium on Networked Systems Design and Implementation (NSDI 25)},
  pages={559--578},
  year={2025}
}

@inproceedings{splitwise,
  title={Splitwise: Efficient generative llm inference using phase splitting},
  author={Patel, Pratyush and Choukse, Esha and Zhang, Chaojie and Shah, Aashaka and Goiri, {\'I}{\~n}igo and Maleki, Saeed and Bianchini, Ricardo},
  booktitle={2024 ACM/IEEE 51st Annual International Symposium on Computer Architecture (ISCA)},
  pages={118--132},
  year={2024},
  organization={IEEE}
}

@article{qwen-vl,
  title={Qwen2-vl: Enhancing vision-language model's perception of the world at any resolution},
  author={Wang, Peng and Bai, Shuai and Tan, Sinan and Wang, Shijie and Fan, Zhihao and Bai, Jinze and Chen, Keqin and Liu, Xuejing and Wang, Jialin and Ge, Wenbin and others},
  journal={arXiv preprint arXiv:2409.12191},
  year={2024}
}

@article{qwen2.5-vl,
  title={Qwen2.5-vl technical report},
  author={Bai, Shuai and Chen, Keqin and Liu, Xuejing and Wang, Jialin and Ge, Wenbin and Song, Sibo and Dang, Kai and Wang, Peng and Wang, Shijie and Tang, Jun and others},
  journal={arXiv preprint arXiv:2502.13923},
  year={2025}
}

@article{qwen2,
  title={Qwen2. 5-vl technical report},
  author={Bai, Shuai and Chen, Keqin and Liu, Xuejing and Wang, Jialin and Ge, Wenbin and Song, Sibo and Dang, Kai and Wang, Peng and Wang, Shijie and Tang, Jun and others},
  journal={arXiv preprint arXiv:2502.13923},
  year={2025}
}

@article{minicpm,
  title={Minicpm: Unveiling the potential of small language models with scalable training strategies},
  author={Hu, Shengding and Tu, Yuge and Han, Xu and He, Chaoqun and Cui, Ganqu and Long, Xiang and Zheng, Zhi and Fang, Yewei and Huang, Yuxiang and Zhao, Weilin and others},
  journal={arXiv preprint arXiv:2404.06395},
  year={2024}
}

@article{deepseek-vl,
  title={Deepseek-vl: towards real-world vision-language understanding},
  author={Lu, Haoyu and Liu, Wen and Zhang, Bo and Wang, Bingxuan and Dong, Kai and Liu, Bo and Sun, Jingxiang and Ren, Tongzheng and Li, Zhuoshu and Yang, Hao and others},
  journal={arXiv preprint arXiv:2403.05525},
  year={2024}
}

@article{mllm-survey,
  title={A survey on multimodal large language models},
  author={Yin, Shukang and Fu, Chaoyou and Zhao, Sirui and Li, Ke and Sun, Xing and Xu, Tong and Chen, Enhong},
  journal={National Science Review},
  volume={11},
  number={12},
  pages={nwae403},
  year={2024},
  publisher={Oxford University Press}
}

@article{llava,
  title={Visual instruction tuning},
  author={Liu, Haotian and Li, Chunyuan and Wu, Qingyang and Lee, Yong Jae},
  journal={Advances in neural information processing systems},
  volume={36},
  pages={34892--34916},
  year={2023}
}

@article{failover-train,
  title={Robust LLM Training Infrastructure at ByteDance},
  author={Wan, Borui and Liu, Gaohong and Song, Zuquan and Wang, Jun and Zhang, Yun and Sheng, Guangming and Wang, Shuguang and Wei, Houmin and Wang, Chenyuan and Lou, Weiqiang and others},
  journal={arXiv preprint arXiv:2509.16293},
  year={2025}
}

@misc{hw-recovery,
  title={Ascend cluster infra recovery},
  year={2025},
  howpublished={\url{https://gitcode.com/ascend-tribe/ascend-cluster-infra/blob/main/HighAvailability/ascend-cluster-infra-infer-recovery.md}}
}

@inproceedings{nanoflow,
  title={$\{$NanoFlow$\}$: Towards Optimal Large Language Model Serving Throughput},
  author={Zhu, Kan and Gao, Yufei and Zhao, Yilong and Zhao, Liangyu and Zuo, Gefei and Gu, Yile and Xie, Dedong and Ye, Zihao and Kamahori, Keisuke and Lin, Chien-Yu and others},
  booktitle={19th USENIX Symposium on Operating Systems Design and Implementation (OSDI 25)},
  pages={749--765},
  year={2025}
}

@article{tokenweave,
  title={TokenWeave: Efficient Compute-Communication Overlap for Distributed LLM Inference},
  author={Gond, Raja and Kwatra, Nipun and Ramjee, Ramachandran},
  journal={arXiv preprint arXiv:2505.11329},
  year={2025}
}

@article{iso,
  title={ISO: Overlap of Computation and Communication within Seqenence For LLM Inference},
  author={Xiao, Bin and Su, Lei},
  journal={arXiv preprint arXiv:2409.11155},
  year={2024}
}

@misc{ascend-docs,
  title={Build models based on capture methods},
  year={2025},
  howpublished={\url{https://www.hiascend.com/document/detail/zh/CANNCommunityEdition/83RC1alpha002/appdevg/acldevg/aclcppdevg_000519.html}}
}

@misc{pytorch-cudagraph,
  title={Accelerating pytorch with CUDA Graphs},
  author={Vinh Nguyen and Michael Carilli and Sukru Burc Eryilmaz and Vartika Singh and Michelle Lin and Natalia Gimelshein and Alban Desmaison and Edward Yang},
  year={2021},
  howpublished={\url{https://pytorch.org/blog/
accelerating-pytorch-with-cuda-graphs/}}
}

@misc{nvidia-cudagraph,
  title={Getting started with CUDA Graphs},
  author={Alan Gray},
  year={2019},
  howpublished={\url{https://developer.nvidia.com/blog/cuda-graphs/}}
}

@inproceedings{vattention,
  title={vattention: Dynamic memory management for serving llms without pagedattention},
  author={Prabhu, Ramya and Nayak, Ajay and Mohan, Jayashree and Ramjee, Ramachandran and Panwar, Ashish},
  booktitle={Proceedings of the 30th ACM International Conference on Architectural Support for Programming Languages and Operating Systems, Volume 1},
  pages={1133--1150},
  year={2025}
}

@article{vtensor,
  title={vtensor: Flexible virtual tensor management for efficient llm serving},
  author={Xu, Jiale and Zhang, Rui and Guo, Cong and Hu, Weiming and Liu, Zihan and Wu, Feiyang and Feng, Yu and Sun, Shixuan and Shao, Changxu and Guo, Yuhong and others},
  journal={arXiv preprint arXiv:2407.15309},
  year={2024}
}

@article{ellm,
  title={eLLM: Elastic Memory Management Framework for Efficient LLM Serving},
  author={Xu, Jiale and Zhang, Rui and Xiong, Yi and Guo, Cong and Liu, Zihan and Zhou, Yangjie and Hu, Weiming and Wu, Hao and Shao, Changxu and Wang, Ziqing and others},
  journal={arXiv preprint arXiv:2506.15155},
  year={2025}
}

@article{prism,
  title={Prism: Unleashing GPU Sharing for Cost-Efficient Multi-LLM Serving},
  author={Yu, Shan and Xing, Jiarong and Qiao, Yifan and Ma, Mingyuan and Li, Yangmin and Wang, Yang and Yang, Shuo and Xie, Zhiqiang and Cao, Shiyi and Bao, Ke and others},
  journal={arXiv preprint arXiv:2505.04021},
  year={2025}
}

@misc{hw-mtp,
  title={Ascend inference cluster},
  year={2025},
  howpublished={\url{https://gitcode.com/ascend-tribe/ascend-inference-cluster/tree/main}}
}

@article{fastmtp,
  title={FastMTP: Accelerating LLM Inference with Enhanced Multi-Token Prediction},
  author={Cai, Yuxuan and Liang, Xiaozhuan and Wang, Xinghua and Ma, Jin and Liang, Haijin and Luo, Jinwen and Zuo, Xinyu and Duan, Lisheng and Yin, Yuyang and Chen, Xi},
  journal={arXiv preprint arXiv:2509.18362},
  year={2025}
}

@article{eagle,
  title={Eagle: Speculative sampling requires rethinking feature uncertainty},
  author={Li, Yuhui and Wei, Fangyun and Zhang, Chao and Zhang, Hongyang},
  journal={arXiv preprint arXiv:2401.15077},
  year={2024}
}

@article{mla,
  title={Transmla: Multi-head latent attention is all you need},
  author={Meng, Fanxu and Tang, Pingzhi and Tang, Xiaojuan and Yao, Zengwei and Sun, Xing and Zhang, Muhan},
  journal={arXiv preprint arXiv:2502.07864},
  year={2025}
}

@article{ascend-eplb,
  title={EfficientMoE: Optimizing Mixture-of-Experts Model Training With Adaptive Load Balance},
  author={Zeng, Yan and Huang, Chengchuang and Mei, Yipeng and Zhang, Lifu and Su, Teng and Ye, Wei and Shi, Wenqi and Wang, Shengnan},
  journal={IEEE Transactions on Parallel and Distributed Systems},
  year={2025},
  publisher={IEEE}
}

@article{xdeepserve,
  title={xDeepServe: Model-as-a-Service on Huawei CloudMatrix384},
  author={Xiao, Ao and He, Bangzheng and Zhang, Baoquan and Huai, Baoxing and Wang, Bingji and Wang, Bo and Xu, Bo and Hou, Boyi and Yang, Chan and Liu, Changhong and others},
  journal={arXiv preprint arXiv:2508.02520},
  year={2025}
}

@article{step-3,
  title={Step-3 is Large yet Affordable: Model-system Co-design for Cost-effective Decoding},
  author={Wang, Bin and Wang, Bojun and Wan, Changyi and Huang, Guanzhe and Hu, Hanpeng and Jia, Haonan and Nie, Hao and Li, Mingliang and Chen, Nuo and Chen, Siyu and others},
  journal={arXiv preprint arXiv:2507.19427},
  year={2025}
}

@article{tree-attention,
  title={Efficient beam search for large language models using Trie-based decoding},
  author={Chan, Brian J and Cheng, Jui-Hung and Huang, Mao Xun and Chen, Chao-Ting and Huang, Hen-Hsen},
  journal={arXiv preprint arXiv:2502.00085},
  year={2025}
}

@article{onerec,
  title={OneRec Technical Report},
  author={Zhou, Guorui and Deng, Jiaxin and Zhang, Jinghao and Cai, Kuo and Ren, Lejian and Luo, Qiang and Wang, Qianqian and Hu, Qigen and Huang, Rui and Wang, Shiyao and others},
  journal={arXiv preprint arXiv:2506.13695},
  year={2025}
}

@article{mtgr,
  title={MTGR: Industrial-Scale Generative Recommendation Framework in Meituan},
  author={Han, Ruidong and Yin, Bin and Chen, Shangyu and Jiang, He and Jiang, Fei and Li, Xiang and Ma, Chi and Huang, Mincong and Li, Xiaoguang and Jing, Chunzhen and others},
  journal={arXiv preprint arXiv:2505.18654},
  year={2025}
}

@article{sparse-meets-dense,
  title={Sparse meets dense: Unified generative recommendations with cascaded sparse-dense representations},
  author={Yang, Yuhao and Ji, Zhi and Li, Zhaopeng and Li, Yi and Mo, Zhonglin and Ding, Yue and Chen, Kai and Zhang, Zijian and Li, Jie and Li, Shuanglong and others},
  journal={arXiv preprint arXiv:2503.02453},
  year={2025}
}

@article{gr-llms,
  title={GR-LLMs: Recent Advances in Generative Recommendation Based on Large Language Models},
  author={Yang, Zhen and Lin, Haitao and Zhang, Ziji and others},
  journal={arXiv preprint arXiv:2507.06507},
  year={2025}
}

@article{pinrec,
  title={PinRec: Outcome-Conditioned, Multi-Token Generative Retrieval for Industry-Scale Recommendation Systems},
  author={Badrinath, Anirudhan and Agarwal, Prabhat and Bhasin, Laksh and Yang, Jaewon and Xu, Jiajing and Rosenberg, Charles},
  journal={arXiv preprint arXiv:2504.10507},
  year={2025}
}

@article{hllm,
  title={Hllm: Enhancing sequential recommendations via hierarchical large language models for item and user modeling},
  author={Chen, Junyi and Chi, Lu and Peng, Bingyue and Yuan, Zehuan},
  journal={arXiv preprint arXiv:2409.12740},
  year={2024}
}

@article{tiger,
  title={Recommender systems with generative retrieval},
  author={Rajput, Shashank and Mehta, Nikhil and Singh, Anima and Hulikal Keshavan, Raghunandan and Vu, Trung and Heldt, Lukasz and Hong, Lichan and Tay, Yi and Tran, Vinh and Samost, Jonah and others},
  journal={Advances in Neural Information Processing Systems},
  volume={36},
  pages={10299--10315},
  year={2023}
}

@article{flashrecovery,
  title={FlashRecovery: Fast and Low-Cost Recovery from Failures for Large-Scale Training of LLMs},
  author={Zhang, Haijun and Wang, Jinxiang and Yu, Zhenhua and Zhang, Yanyong and Ji, Xuejie and Mao, Kaining and Zhang, Jun and Zhang, Yaqing and Wu, Ting and Jie, Fei and others},
  journal={arXiv preprint arXiv:2509.03047},
  year={2025}
}

@article{duan2024efficient,
  title={Efficient training of large language models on distributed infrastructures: a survey},
  author={Duan, Jiangfei and Zhang, Shuo and Wang, Zerui and Jiang, Lijuan and Qu, Wenwen and Hu, Qinghao and Wang, Guoteng and Weng, Qizhen and Yan, Hang and Zhang, Xingcheng and others},
  journal={arXiv preprint arXiv:2407.20018},
  year={2024}
}

@article{wu2023transom,
  title={Transom: An efficient fault-tolerant system for training llms},
  author={Wu, Baodong and Xia, Lei and Li, Qingping and Li, Kangyu and Chen, Xu and Guo, Yongqiang and Xiang, Tieyao and Chen, Yuheng and Li, Shigang},
  journal={arXiv preprint arXiv:2310.10046},
  year={2023}
}

@inproceedings{mohan2021checkfreq,
  title={$\{$CheckFreq$\}$: Frequent,$\{$Fine-Grained$\}$$\{$DNN$\}$ Checkpointing},
  author={Mohan, Jayashree and Phanishayee, Amar and Chidambaram, Vijay},
  booktitle={19th USENIX Conference on File and Storage Technologies (FAST 21)},
  pages={203--216},
  year={2021}
}

@inproceedings{chen2023cost,
  title={A cost-efficient failure-tolerant scheme for distributed dnn training},
  author={Chen, Menglei and Hua, Yu and Bai, Rong and Huang, Jianming},
  booktitle={2023 IEEE 41st International Conference on Computer Design (ICCD)},
  pages={150--157},
  year={2023},
  organization={IEEE}
}

@article{singh2024efficiently,
  title={Efficiently Serving Large Multimodal Models Using EPD Disaggregation},
  author={Singh, Gursimran and Wang, Xinglu and Hu, Yifan and Yu, Timothy and Xing, Linzi and Jiang, Wei and Wang, Zhefeng and Bai, Xiaolong and Li, Yi and Xiong, Ying and others},
  journal={arXiv preprint arXiv:2501.05460},
  year={2024}
}

@article{ning2024inf,
  title={Inf-MLLM: Efficient streaming inference of multimodal large language models on a single GPU},
  author={Ning, Zhenyu and Zhao, Jieru and Jin, Qihao and Ding, Wenchao and Guo, Minyi},
  journal={arXiv preprint arXiv:2409.09086},
  year={2024}
}

@misc{claude,
  title={Anthropic Claude},
  author={{Anthropic}},
  year={2023},
  howpublished={\url{https://claude.ai/}}
}

@article{deepseek-r1,
  title={Deepseek-r1: Incentivizing reasoning capability in llms via reinforcement learning},
  author={Guo, Daya and Yang, Dejian and Zhang, Haowei and Song, Junxiao and Zhang, Ruoyu and Xu, Runxin and Zhu, Qihao and Ma, Shirong and Wang, Peiyi and Bi, Xiao and others},
  journal={arXiv preprint arXiv:2501.12948},
  year={2025}
}

@misc{copilot,
  title={Github Copilot},
  author={{Github}},
  year={2022},
  howpublished={\url{https://github.com/features/copilot}}
}

@article{qwen2.5-coder,
  title={Qwen2. 5-coder technical report},
  author={Hui, Binyuan and Yang, Jian and Cui, Zeyu and Yang, Jiaxi and Liu, Dayiheng and Zhang, Lei and Liu, Tianyu and Zhang, Jiajun and Yu, Bowen and Lu, Keming and others},
  journal={arXiv preprint arXiv:2409.12186},
  year={2024}
}

@article{starcoder,
  title={Starcoder: may the source be with you!},
  author={Li, Raymond and Allal, Loubna Ben and Zi, Yangtian and Muennighoff, Niklas and Kocetkov, Denis and Mou, Chenghao and Marone, Marc and Akiki, Christopher and Li, Jia and Chim, Jenny and others},
  journal={arXiv preprint arXiv:2305.06161},
  year={2023}
}

@article{bowen2009document,
  title={Document analysis as a qualitative research method},
  author={Bowen, Glenn A},
  journal={Qualitative research journal},
  volume={9},
  number={2},
  pages={27--40},
  year={2009},
  publisher={Emerald Group Publishing Limited}
}

@article{karim2025transforming,
  title={Transforming Data Annotation with AI Agents: A Review of Architectures, Reasoning, Applications, and Impact},
  author={Karim, Md Monjurul and Khan, Sangeen and Van, Dong Hoang and Liu, Xinyue and Wang, Chunhui and Qu, Qiang},
  journal={Future Internet},
  volume={17},
  number={8},
  pages={353},
  year={2025},
  publisher={MDPI}
}

@article{auto-scaling,
  title={A review of auto-scaling techniques for elastic applications in cloud environments},
  author={Lorido-Botran, Tania and Miguel-Alonso, Jose and Lozano, Jose A},
  journal={Journal of grid computing},
  volume={12},
  number={4},
  pages={559--592},
  year={2014},
  publisher={Springer}
}

@article{sun2025hygen,
  title={HyGen: Efficient LLM Serving via Elastic Online-Offline Request Co-location},
  author={Sun, Ting and Wang, Penghan and Lai, Fan},
  journal={arXiv preprint arXiv:2501.14808},
  year={2025}
}

@article{wang2025echo,
  title={Echo: Efficient Co-Scheduling of Hybrid Online-Offline Tasks for Large Language Model Serving},
  author={Wang, Zhibin and Li, Shipeng and Li, Xue and Zhou, Yuhang and Zhang, Zhonghui and Wang, Zibo and Gu, Rong and Tian, Chen and Yang, Kun and Zhong, Sheng},
  journal={arXiv preprint arXiv:2504.03651},
  year={2025}
}

@article{borui2025efficient,
  title={Efficient LLM Serving on Hybrid Real-time and Best-effort Requests},
  author={Borui, Wan and Juntao, Zhao and Chenyu, Jiang and Chuanxiong, Guo and Chuan, Wu},
  journal={arXiv preprint arXiv:2504.09590},
  year={2025}
}

@article{zeng2025efficientmoe,
  title={EfficientMoE: Optimizing Mixture-of-Experts Model Training With Adaptive Load Balance},
  author={Zeng, Yan and Huang, Chengchuang and Mei, Yipeng and Zhang, Lifu and Su, Teng and Ye, Wei and Shi, Wenqi and Wang, Shengnan},
  journal={IEEE Transactions on Parallel and Distributed Systems},
  year={2025},
  publisher={IEEE}
}

@article{xue2024moe,
  title={Moe-infinity: Offloading-efficient moe model serving},
  author={Xue, Leyang and Fu, Yao and Lu, Zhan and Mai, Luo and Marina, Mahesh},
  journal={arXiv preprint arXiv:2401.14361},
  year={2024}
}

@article{fedus2022switch,
  title={Switch transformers: Scaling to trillion parameter models with simple and efficient sparsity},
  author={Fedus, William and Zoph, Barret and Shazeer, Noam},
  journal={Journal of Machine Learning Research},
  volume={23},
  number={120},
  pages={1--39},
  year={2022}
}

@article{dai2024deepseekmoe,
  title={Deepseekmoe: Towards ultimate expert specialization in mixture-of-experts language models},
  author={Dai, Damai and Deng, Chengqi and Zhao, Chenggang and Xu, RX and Gao, Huazuo and Chen, Deli and Li, Jiashi and Zeng, Wangding and Yu, Xingkai and Wu, Yu and others},
  journal={arXiv preprint arXiv:2401.06066},
  year={2024}
}

@article{qwen3,
  title={Qwen3 technical report},
  author={Yang, An and Li, Anfeng and Yang, Baosong and Zhang, Beichen and Hui, Binyuan and Zheng, Bo and Yu, Bowen and Gao, Chang and Huang, Chengen and Lv, Chenxu and others},
  journal={arXiv preprint arXiv:2505.09388},
  year={2025}
}

@misc{qwen25,
      title={Qwen2.5 Technical Report}, 
      author={Qwen Team},
      year={2025},
      eprint={2412.15115},
      archivePrefix={arXiv},
      primaryClass={cs.CL},
      url={https://arxiv.org/abs/2412.15115}, 
}

@article{ecoserve,
  title={EcoServe: Enabling Cost-effective LLM Serving with Proactive Intra-and Inter-Instance Orchestration},
  author={Du, Jiangsu and Zhang, Hongbin and Wei, Taosheng and Zheng, Zhenyi and Wu, Kaiyi and Chen, Zhiguang and Lu, Yutong},
  journal={arXiv preprint arXiv:2504.18154},
  year={2025}
}

}
\newpage

\end{document}